\documentclass[aps,prb,reprint,twocolumn,tightenlines,citeautoscript,%
showpacs,floatfix,superscriptaddress]{revtex4}
\usepackage{bm}
\usepackage{amssymb}
\usepackage{graphicx}
\usepackage{amsmath}
\usepackage{textcomp}
\usepackage{lscape}
\usepackage{bigstrut}

\begin{document}
\title{Two-photon Indirect Optical Injection and Two-color Coherent Control in
  Bulk Silicon}
\author{J. L. Cheng}
\affiliation{Department of Physics and Institute for Optical Sciences,
University of Toronto, 60 St. George Street, Toronto, Ontario, Canada
M5S 1A7}
\author{J. Rioux}
\thanks{Current address: Department of Physics, University of
  Konstanz, D-78457 Konstanz, Germany.}
\affiliation{Department of Physics and Institute for Optical Sciences,
University of Toronto, 60 St. George Street, Toronto, Ontario, Canada
M5S 1A7}
\author{J. E. Sipe}%
\affiliation{Department of Physics and Institute for Optical Sciences,
University of Toronto, 60 St. George Street, Toronto, Ontario, Canada
M5S 1A7}

\date{\today}
\begin{abstract}
  Using an empirical pseudopotential description of electron states
  and an adiabatic bond charge model for phonon states in bulk silicon, we
  theoretically investigate two-photon indirect optical
  injection of carriers and spins and two-color
  coherent control of the motion of the injected carriers and
  spins. For two-photon indirect carrier and spin injection,
  we identify the selection rules of band edge transitions, the
  injection in each conduction band valley, and the injection from
  each phonon branch at 4~K and 300~K. At 4~K, the
  TA phonon-assisted transitions dominate the injection at low photon energies, and
  the TO phonon-assisted at high photon energies. At 300~K, the
  former dominates at all photon energies of interest. The carrier injection shows
  anisotropy and linear-circular dichroism with respect to light
  propagation direction. For light propagating along the
  $\langle001\rangle$ direction, the carrier injection exhibits valley anisotropy,
  and the injection into the $Z$ conduction band valley is larger than that into the $X/Y$
  valleys. For $\sigma^-$ light propagating along the
  $\langle001\rangle$ ($\langle 111\rangle$) direction, the degree of
  spin polarization gives a maximum value about $20\%$ ($6\%$) at 4~K and 
  $-10\%$ ($20\%$) at 300~K, and at both temperature shows abundant structure near the
  injection edges due to contributions from different phonon branches. For
  the two-color coherent current injection with an incident optical
  field composed of a fundamental frequency and its
  second harmonic, the response tensors of the electron (hole) charge and
  spin currents are calculated at 4~K and 300~K. We show the
  current control for three different polarization scenarios: For
  co-circularly polarized beams, the direction of the charge current and the
  polarization direction of the spin current can be controlled by a
  relative-phase parameter; for the co-linearly and cross-linearly
  polarized beams, the current amplitude can be controlled by that parameter.
  The spectral dependence of the maximum swarm velocity shows that the
  direction of charge current reverses under increase in photon energy.
\end{abstract}


\maketitle

\section{Introduction}
Silicon is a dominant material in the microelectronics industry. It has
also attracted much attention in
optoelectronics\cite{SiliconPhotonics_Lockwood_Pavesi,IEEE_12_1678_2006_Soref,J.LightwaveTech._23_4222_2005_Lipson},
due to its low absorption
at telecommunication wavelengths near $1.55~\mu$m, and in
spintronics\cite{Rev.Mod.Phys._76_323_2004_Zutic,ActaPhys.Slovaca_57_565_2007_Fabian,Phys.Rep._493_61_2010_Wu},
due to its long spin transport length and spin relaxation
time\cite{Nature_447_295_2007_Appelbaum,Phys.Rev.Lett._99_177209_2007_Huang,NaturePhys._3_542_2007_Jonker,Phys.Rev.B_2_2429_1970_Lepine,Phys.Rev.B_78_054446_2008_Mavropoulos,Phys.Rev.B_79_075303_2009_Zhang,Phys.Rev.Lett._105_037204_2010_Li}. In
both fields, a full understanding of the optical properties in bulk
silicon is very important for further applications. Optical methods can
provide an effective way to generate carriers and
spins in semiconductors, to
control\cite{Phys.StatusSolidiB_243_2278_2006_vanDriel,CoherentControlOfPhotocurrentsinSemiconductors_Driel,CoherentControl_Driel}
their motions by the phase coherence of different components of
incident laser beams, and to detect the properties of carriers and
spins\cite{optical_orientation}. Recently, the direct detection of
spin currents using second-order nonlinear optical effects has been
proposed\cite{Phys.Rev.Lett._104_256601_2010_Wang} and realized
experimentally\cite{Nat.Phys._6_875_2010_Werake}. 

Because silicon is an indirect gap semiconductor, with an indirect gap
$E_{ig}=1.17$~eV and a direct gap $E_g=3.4$~eV \cite{landolt_Si},
there is a degenerate indirect ``$\ell$''-photon optical transition
assisted by phonon emission or absorption at
$\ell\hbar\omega<E_g$. This optical response is about two
orders of magnitude weaker than that in direct gap
semiconductors. While the weak response results in low loss, which is
important in realizing optoelectronics devices, it can make
optical coherent control less effective.

By using circularly
polarized light, spin polarized carriers can be injected\cite{optical_orientation}. Generally, one- and two-photon injection are the
most widely used schemes. For coherent current control, the minimum requirements depend on the
semiconductor crystal structures: For low symmetry semiconductor
structures with nonvanishing second order nonlinearity, such as the
wurtzite structure\cite{Appl.Phys.Lett._75_2581_1999_Laman}, current
can be injected by even a single frequency
laser beam; for high symmetry
semiconductor structures with vanishing second order nonlinearity but
non-vanishing third-order nonlinearity, such as the diamond structures,
current injection requires at least a two-color laser pulse with one
fundamental frequency and its $\ell^{th}$ harmonic (``$1+\ell$''
effects). The control parameters are taken as a relative-phase parameter
between Cartesian components  or between the frequency components of
the two-color laser beams. However, most coherent control studies to
date have focused on absorption across the direct gap\cite{Appl.Phys.Lett._95_092107_2009_Loren,Phys.Rev.Lett._85_5432_2000_Bhat,Phys.Rev.B_68_165348_2003_Najmaie}, even when
considering the indirect gap semiconductors\cite{Phys.Rev.B_81_155215_2010_Rioux}; seldom has coherent
control by absorption across an indirect gap been considered\cite{Appl.Phys.Lett._97_212106_2010_Zhao,NaturePhys._3_632_2007_Costa,Phys.Rev.B_77_085201_2008_Spasenovic}, due to the weak optical
response. For silicon, which has diamond structure and vanishing
second-order nonlinearity, it is only the second of the coherent
control schemes mentioned above that is applicable.

For two-photon indirect optical carrier injection in bulk silicon,
most experimental studies have focused on the two-photon absorption
coefficient\cite{Phys.Rev.Lett._30_901_1973_Reintjes,Appl.Phys.Lett._80_416_2002_Tsang,Appl.Phys.Lett._82_2954_2003_Dinu,Appl.Phys.Lett._90_191104_2007_Bristow,Appl.Phys.Lett._91_071113_2007_Zhang,Appl.Phys.Lett._91_021111_2007_Lin}
and its anisotropy\cite{Appl.Phys.Lett._91_071113_2007_Zhang}, which is important in optoelectronics
devices; theoretical
studies\cite{IEEEJ.QuantumElectron._39_1398_2003_Dinu,J.Phys.B_39_2737_2006_Garcia,Phys.Stat.Solidi(b)_184_519_1994_Hassan,Phys.Stat.Solidi(b)_186_303_1994_Hassan}
are mostly based on the parabolic band approximation and on a
phenomenological electron-phonon interaction. 
For the current injection by coherent control, Costa {\it et
  al.}\cite{NaturePhys._3_632_2007_Costa} and Spasenovi\'c {\it et
  al.} \cite{Phys.Rev.B_77_085201_2008_Spasenovic} used THz radiation to detect ``1+2'' injected current in bulk silicon, and
confirmed that the current can be controlled by the phase parameter
of the laser beams. Zhao and Smirl
\cite{Appl.Phys.Lett._97_212106_2010_Zhao} measured the time- and
space-evolution of the indirect optical injected electrons and holes by 
phase-dependent differential transmission techniques. Yet a full band
structure investigations of the two-photon indirect optical injection of spins
and spin current are still lacking.

Previously we studied the
one-photon indirect optical injection of carriers and
spins\cite{Phys.Rev.B_83_165211_2011_Cheng}, and the spectral
dependence of the two-photon
indirect absorption coefficients and their phonon-resolved injection
rates at 4~K and 300~K\cite{Appl.Phys.Lett._98_131101_2011_Cheng}. In this 
paper, we continue the study of the two-photon indirect optical injection of
carriers and spins, and consider as well the coherent control of the
injected charge and spin currents by ``1+2'' effects.
We present the detailed results of two-photon indirect
carrier and spin injection under $\sigma^-$ light propagating along
$\langle 001\rangle$ and $\langle111\rangle$
directions; due to the symmetries of bulk silicon, the
  injection with $\sigma^+$ light has the same carrier and spin
  injection as with $\sigma^-$ light, but with the opposite spin polarization. The injection in each conduction
band valley, the anisotropy and the linear-circular dichroism with
respect to light propagation direction, the corresponding phonon-resolved
spectra, and the degree of spin polarization (DSP) are discussed. We
also consider the coherent control of the motion of optically injected
electrons and holes under particular two-color optical fields: co-circularly
polarized beams, co-linearly polarized beams, and cross-linearly
polarized beams. 

In optical absorption, the electron-hole interaction plays an
important role especially in determining the correct absorption 
edges. First principle
studies\cite{Phys.Rev.B_62_4927_2000_Rohlfing} of the direct gap
optical absorption shows that the excitonic effect can strongly change
the lineshape even for high photon energies  in silicon. For 
indirect one-\cite{Phys.Rev._108_1384_1957_Elliott}  and
two-photon injection\cite{Phys.Stat.Solidi(b)_184_519_1994_Hassan,Phys.Stat.Solidi(b)_186_303_1994_Hassan},
investigations within the parabolic band approximation show that this
neglect does not change the absorption lineshapes at energies more than
a few binding energies above the band gap; however, a full band structure
investigation is still lacking due to difficulty in numerical
calculation of the wave functions of the electron-hole pair. In this
paper, as a preliminary calculation, we neglect the excitonic effect.

We organize the paper as follows. Two-photon indirect carrier and
spin injection are presented in Sec.~\ref{sec:carrierspininjection}. In this section, we first
describe a perturbation model for two-photon indirect optical injection, and
then give the numerical results under an empirical pseudopotential
model for electronic states and an adiabatic bond charge model for phonon
states. In Sec.~\ref{sec:coherentcontrol}, we study the interference
current injection under a two-color laser beam and the coherent
control.  We conclude in Sec.~\ref{sec:conclusion}.

\section{\label{sec:carrierspininjection}Two-photon indirect Carrier and spin injection}
\subsection{Model for two-photon indirect injection}
For an incident laser beam with electric field ${\bm E}(t) = {\bm
  E}_\omega e^{-i\omega t} +  c.c$, the two-photon optical injection
rates of electrons and their spins are generally written as 
\begin{eqnarray}
  \dot{n} &=& \xi^{abcd}E_\omega^aE_\omega^b
  (E_\omega^cE_\omega^d)^{\ast}\ ,\nonumber\\
  \dot{S}^f
  &=&\zeta^{fabcd}E_\omega^aE_\omega^b(E_\omega^cE_\omega^d)^{\ast}\ ,
\label{eq:rates}
\end{eqnarray}
From these rates, the actual injected carrier density and spin density
can be calculated once the pulse duration is specified.
In this paper, superscripts indicate Cartesian coordinates, and
repeated superscripts are to be summed over. For bulk silicon, the
lowest conduction band has six equivalent valleys, which are usually
denoted as $X, \bar{X}, Y, \bar{Y}, Z, \bar{Z}$. The two-photon
  indirect transitions have the same initial and final states as
that of one-photon indirect transitions\cite{Phys.Rev.B_83_165211_2011_Cheng}. The injection
coefficients can be written as the form $ {\cal
  A}^{abcd}=\sum_{I}{\cal  A}^{abcd}_I$ with ${\cal A}^{abcd}_I$
identifying the injection into the $I^{th}$ valley. Fermi's golden rule gives
${\cal A}^{abcd}_I =\sum_{cv\lambda\pm}{\cal
  A}^{abcd}_{I;cv\lambda\pm}$ with 
\begin{eqnarray}
  {\cal A}^{abcd}_{I;cv\lambda\pm} &=& \frac{2\pi}{\hbar}\sum_{\bm
    k_c\in I, \bm k_v}\delta(\varepsilon_{c\bm
  k_c}-\varepsilon_{v\bm k_v}\pm\hbar\Omega_{(\bm
  k_c-\bm k_v)\lambda}-2\hbar\omega) \nonumber\\
  &\times&N_{(\bm k_c-\bm
      k_v)\lambda\pm} {\cal
    A}^{abcd}_{c\bm k_cv\bm k_v\lambda}\ ,\label{eq:X}
\end{eqnarray}
\begin{equation}
{\cal A}^{abcd}_{c\bm k_cv\bm k_v\lambda} = \sum_{\sigma_c\sigma_c^{\prime}\sigma_v}\langle
\bar{c}^{\prime}\bm k_c|\hat{\cal A}|\bar{c}\bm
  k_c\rangle W^{ab}_{\bar{c}\bm k_c\bar{v}\bm 
    k_v\lambda}[W^{cd}_{\bar{c}^{\prime}\bm k_c\bar{v}\bm
      k_v\lambda}]^{\ast}\ .
\label{eq:calA}
\end{equation}
The coefficient ${\cal A}^{abcd}_{I;cv\lambda\pm}$ gives the
contribution to the injection by indirect optical transition between conduction
band $c$ and valence band $v$, with the assistance of an emitted ($+$)
or absorbed ($-$) phonon in the $\lambda^{th}$-mode; there are two
modes each for the
transverse acoustic (TA) and optical (TO) branches, and one mode each for the longitudinal acoustic (LA) and
optical (LO) branches.  The operator $\hat{A}$ in Eq.~(\ref{eq:calA})
stands for the identity operator in carrier injection, and the
$f^{th}$ component of spin operator in spin injection. The
optical transition matrix elements are given as
\begin{eqnarray}
  W^{ab}_{\bar{c}\bm k_c\bar{v}\bm k_v\lambda} &=& \frac{1}{2}
  \left(\frac{e}{\hbar\omega}\right)^2
  \sum_{\bar{n}\bar{m}}\bigg[\frac{M_{\bar{c}\bm k_c\bar{n}\bm k_v;\lambda}v^a_{\bar{n}\bar{m}\bm k_v}v^b_{\bar{m}\bar{v}\mathbf
      k_v}}{(\omega_{nv\bm k_v}-2\omega)(\omega_{mv\mathbf
      k_v}-\omega)}\nonumber\\
&-&\frac{v^a_{\bar{c}\bar{n}\bm k_c}M_{\bar{n}\bm k_c\bar{m}\bm k_v;\lambda}v^b_{\bar{m}\bar{v}\bm
    k_v}}{(\omega_{cn\bm k_c}-\omega)(\omega_{mv\bm
    k_v}-\omega)}\nonumber\\
&+&\frac{v^a_{\bar{c}\bar{n}\bm k_c}v^b_{\bar{n}\bar{m}\bm k_c}M_{\bar{m}\bm k_c\bar{v}\bm
    k_v;\lambda}}{(\omega_{cn\bm k_c}-\omega)(\omega_{cm\bm
    k_c}-2\omega)}\bigg] + \{a\leftrightarrow b\}
\label{eq:W}
\end{eqnarray}
where $e=|e|$. Here $\bm k_c$ and $\bm k_v$ are the electron and hole wave vectors
respectively; $\bar{c}=\{c,\sigma_c\}$,
$\bar{c}^{\prime}=\{c,\sigma_c^{\prime}\}$, and
$\bar{v}=\{v,\sigma_v\}$ are full band indexes with $\sigma_c$,
$\sigma_c^{\prime}$ and $\sigma_v$ being the spin indexes; $\bar{n}$
and $\bar{m}$ are band indices for intermediate states; $|\bar{c}\bm
k_c\rangle$ and $\varepsilon_{c\bm k_c}$ are the electron states and
its energy respectively; and $\omega_{nm}(\bm k)$ is defined as
$\hbar\omega_{nm}(\bm k)=\varepsilon_{n\bm 
  k}-\varepsilon_{m\bm k}$. The phonon energy is given by
$\hbar\Omega_{\bm q\lambda}$ for wavevector $\bm q$ and mode
$\lambda$, the equilibrium phonon
number is $N_{\bm q\lambda}$, and $N_{\bm q\lambda\pm}=N_{\bm
  q\lambda}+\frac{1}{2}\pm \frac{1}{2}$. The velocity matrix elements are $\bm
v_{\bar{n}\bar{m}}(\bm k)=\langle \bar{n}\bm k|\hat{\bm v}|\bar{m}\bm
k\rangle$ with the velocity operator $\hat{\bm v}=\partial
H_e/\partial \bm p$, and $H_e$ is the unperturbed electron
Hamiltonian. The electron-phonon interaction is written as
$H^{ep}=\sum_{\bm q\lambda}H_{\lambda}^{ep}(\bm q)(a_{\bm
  q\lambda}+a_{-\bm q\lambda}^{\dag})$ with $a_{\bm q\lambda}$
standing for the phonon annihilation operator. Its matrix elements are $M_{\bar{n}\bm
  k_c\bar{m}\bm k_v\lambda}=\langle \bar{n}\bm
k_c|H^{ep}_{\lambda}(\bm k_c-\bm k_v)|\bar{m}\bm k_v\rangle$. 

The injection coefficient $\xi^{abcd}_I$ is a fourth-order tensor and
$\zeta^{fabcd}_I$ is a fifth-order pseudotensor. Both of them are
symmetric on exchange of indices $a$ and $b$, and on exchange of
indices $c$ and $d$. They have the properties
$\left(\xi^{abcd}_I\right)^{\ast}=\xi_I^{cdab}$ and
$\left(\zeta^{fabcd}_I\right)^{\ast}=\zeta_I^{fcdab}$. Furthermore, time-reversal symmetry gives
$\xi_I^{abcd}=\left(\xi_{\bar{I}}^{abcd}\right)^{\ast}$ and $\zeta_I^{fabcd}=-\left(\zeta_{\bar{I}}^{fabcd}\right)^{\ast}$. In bulk silicon,
each conduction band valley has $C_{4v}$ symmetry. Under this
symmetry, $\xi_Z^{abcd}$ has six nonzero independent components,
{
\allowdisplaybreaks
\begin{eqnarray}
  \xi^{xxxx}_Z &=& \xi^{yyyy}_Z\ ,\nonumber\\
  \xi^{xxyy}_{Z} &=& \xi^{yyxx}_{Z}\ ,\nonumber\\
  \xi^{zzxx}_{Z} &=& \xi^{zzyy}_{Z}\ ,\nonumber\\
  \xi^{xyxy}_{Z} &\ ,&\nonumber\\
  \xi^{xzxz}_{Z} &=& \xi^{yzyz}\ ,\nonumber\\
  \xi^{zzzz}_{Z} &\ ; &
\label{eq:nz-xi}
\end{eqnarray}}
$\zeta_Z^{fabcd}$ also has six nonzero independent components,
\begin{eqnarray}
  \zeta_Z^{zxyxx}&=& - \zeta_Z^{zxyyy}\ ,\nonumber\\
  \zeta_Z^{zyzxz}&=& - \zeta_Z^{zxzyz}\ ,\nonumber\\
  \zeta_Z^{xyzyy}&=& - \zeta_Z^{yxzxx}\ ,\nonumber\\
  \zeta_Z^{xyzxx}&=& - \zeta_Z^{yxzyy}\ ,\nonumber\\
  \zeta_Z^{xxzxy}&=& - \zeta_Z^{yyzxy}\ ,\nonumber\\
  \zeta_Z^{xzzyz}&=& - \zeta_Z^{yzzxz}\ .
\label{eq:nz-zeta}
\end{eqnarray}
The injection coefficients ${\cal A}^{abcd}_I$ can be obtained by a
proper rotation operation that transforms the $Z$ valley to the $I^{th}$
valley. Using inversion and time-reversal symmetries, all
$\xi^{abcd}$ are identified as real numbers, and all $\zeta^{fabcd}$
are pure imaginary numbers; ${\cal A}^{abcd}_{I;cv\lambda\pm}$ shares
the same symmetry properties as ${\cal A}^{abcd}_I$, while ${\cal
  A}^{abcd}$ belongs to the higher symmetry group $O_h$, and has fewer nonzero
independent components
\begin{eqnarray}
  && \xi^{xxxx} = \xi^{yyyy} = \xi^{zzzz}\ ,\nonumber\\
  && \xi^{xxyy} = \xi^{xxzz} = \xi^{yyzz} = \xi^{yyxx} = \xi^{zzxx} =
  \xi^{zzyy}\ ,\nonumber\\
  && \xi^{xyxy} = \xi^{xzxz} = \xi^{yzyz}\ ,
\end{eqnarray}
and 
\begin{eqnarray}
  &&  \zeta^{zxyxx} = -\zeta^{zyxyy} = \zeta^{yzxzz} \nonumber\\
  && \quad\quad\quad = -\zeta^{yxzxx} =
  \zeta^{xyzyy} = -\zeta^{xzyzz}\ ,\nonumber \\
  && \zeta^{xxzxy} = -\zeta^{xyxxz} = \zeta^{yxyyz} \nonumber\\
  &&\quad\quad\quad  = -\zeta^{yzyyx} =
  \zeta^{zyzxz} = -\zeta^{zzxyz}\ .
\end{eqnarray}
All these components are related to the nonzero injection coefficients in the $Z$ valley by 
\begin{eqnarray}
  \xi^{xxxx} &=& 4\xi_Z^{xxxx} + 2\xi_Z^{zzzz}\ ,\nonumber\\
  \xi^{xxyy} &=& 4\xi_Z^{zzxx} + 2\xi_Z^{xxyy}\
  ,\nonumber\\
  \xi^{xyxy} &=& 4\xi_Z^{xzxz} + 2\xi_Z^{xyxy}\ ,\nonumber\\
  \zeta^{xxzxy} &=& 4\zeta_Z^{xxzxy} + 2\zeta_Z^{zyzxz} \ ,\nonumber\\
  \zeta^{zxyxx} &=& 2\zeta_Z^{zxyxx} +2\zeta_Z^{xzzyz} +
  2\zeta_Z^{xyzyy}\ .
\label{eq:connect}
\end{eqnarray}
With all these coefficients, the injection rates for
laser pulse with any polarization and propagating directions can be evaluated. In
Appendix \ref{app:injectionrates}, we give in detail the carrier and spin
injection rates for circularly-polarized light with any
propagating direction, and the carrier injection rates for
linearly-polarized light with any polarization and propagating
directions. In the following, we focus on light propagating along
$\langle001\rangle$ and $\langle111\rangle$ directions.

\subsection{Results}
For quantitative calculations of the two-photon indirect injection
rates, a full band-structure description of the electron and phonon states
is necessary. Here we use an empirical pseudopotential
model\cite{Phys.Rev.B_10_5095_1974_Chelikowsky,Phys.Rev.B_14_556_1976_Chelikowsky,Phys.Rev._149_504_1966_Weisz}
for electron states and an adiabatic bond charge model\cite{Phys.Rev.B_15_4789_1977_Weber} for phonon
states. All the parameters used in the empirical pseudopotential
model and the adiabatic bond charge model are the same as those in the
calculation of one-photon optical spin
injection\cite{Phys.Rev.B_83_165211_2011_Cheng}. From the empirical pseudopotential
model, 
the calculated direct band gap is $E_g=3.43$~eV, the indirect band
gap is $E_{ig}=1.17$~eV; the band edge for the conduction band is
located at $\bm k_c^0=0.85~\overrightarrow{\Gamma X}$, and for the
valence bands at the $\Gamma$ point, $\bm k_v^0=0$. From the adiabatic bond charge model,
the energies for phonons with wavevector $\bm k_c^0$ are $19$ (TA),
$43$ (LA), $53$ (LO), and $57$ (TO) meV. Within the pseudopotential
scheme we determine the electron-phonon interaction, and then evaluate
the matrix elements $H_\lambda^{ep}(\bm q)$ using the calculated
electron and phonon wavefunctions. With all these quantities
calculated, the two-photon indirect gap transition matrix elements in
Eq.~(\ref{eq:W}) are calculated using the lowest $30$ electron
bands as intermediate states to ensure convergence. The injection
coefficients given in Eq.~(\ref{eq:calA}) are evaluated using an
improved linear analytic tetrahedral method\cite{Phys.Rev.B_83_165211_2011_Cheng}.    

\begin{figure}[htp]
  \centering
  \includegraphics[width=8cm]{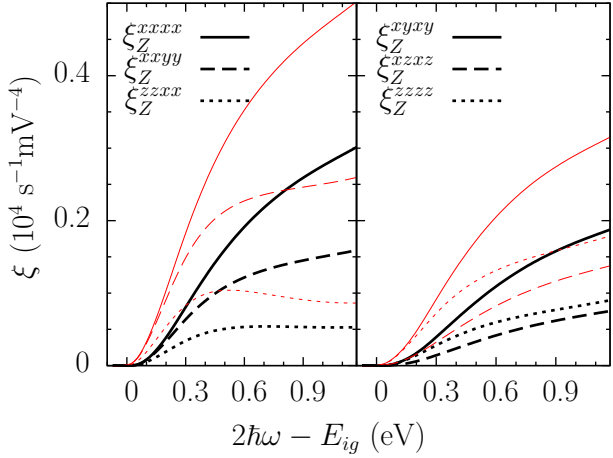}
  \caption{(Color online) Spectra of $\xi_{Z}^{abcd}$
    at 4~K (thick black curves) and 300~K (thin red curves). }
  \label{fig:carriercomponents}
\end{figure}

\begin{figure}[htp]
  \centering
  \includegraphics[width=8cm]{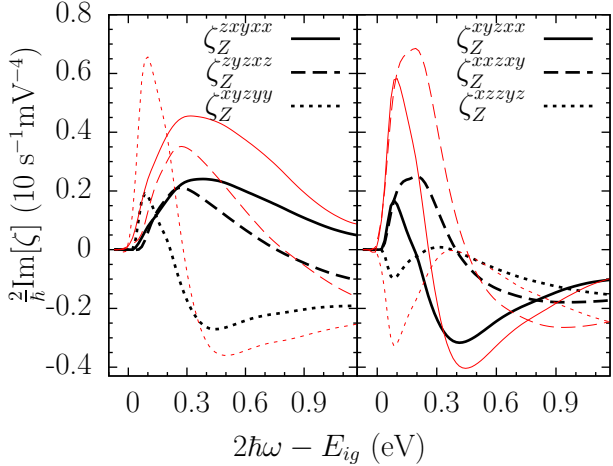}
  \caption{(Color online) Spectra of
    $\frac{2}{\hbar}\text{Im}[\zeta_Z^{fabcd}]$ at 4~K (thick black
    curves) and 300~K (thin red curves).}
  \label{fig:spincomponents}
\end{figure}

In our calculation, the valence bands include heavy hole (HH), light
hole (LH), and spin split-off (SO) bands; the conduction bands include
the lowest two conduction bands. Our results are
shown in Fig.~\ref{fig:carriercomponents} for the spectra of nonzero components
of $\xi^{abcd}_Z$ and in Fig.~\ref{fig:spincomponents} for the spectra
of nonzero components of $\zeta^{fabcd}_Z$ at 4~K and 300~K,
respectively. The full two-photon indirect gap injection rates in Eq.~(\ref{eq:rates}) can be
identified for any polarization of the electric field using Eq.~(\ref{eq:connect}). Comparing
the injection rates given in Eqs.~(\ref{eq:X}) and (\ref{eq:calA}) with the one-photon indirect optical
injection rates\cite{Phys.Rev.B_83_165211_2011_Cheng}, we find
that these two formulas differ only in the transition matrix elements given in
Eq.~(\ref{eq:calA}). Therefore they show similar temperature
dependence, which is mainly determined by the phonon number, and similar
contributions from each valence band, which is mainly determined by the
joint density of states. 

In a previous paper\cite{Appl.Phys.Lett._98_131101_2011_Cheng}, we
discussed in detail the photon energy, temperature, and phonon branch
dependence of the total carrier injection coefficients $\xi^{xxxx}$,
$\xi^{xxyy}$, and $\xi^{xyxy}$. Here the $\xi^{abcd}_Z$ in Fig.~\ref{fig:carriercomponents}
show similar properties: For excess photon energies
$2\hbar\omega-E_{ig}$ of interest, $\xi_Z^{zzxx}$ first increases with increasing photon
energy, and then slightly decreases; all the other components increase
monotonically. In contrast, all the components of $\zeta^{fabcd}_Z$, given in Fig.~\ref{fig:spincomponents},
show a complicated photon energy dependence. All
injection rates at 300~K are larger than those at 4~K due to the
larger phonon populations.

To better understand these results, we first consider the properties of
transitions around the band edges. Then we turn to
the injection rates of carriers and spins, and the DSP under
$\sigma^-$ light propagating along two different directions.

\subsubsection{Transitions near band edges}
One can try to simplify the description of the indirect two-photon injection around the band
edges using the high symmetry at the band edge. We symmetrize the indirect two-photon injection rates as 
\begin{eqnarray}
  {\cal A}^{abcd}_{I;cv\tau\pm} &=& \frac{2\pi}{\hbar}\sum_{\bm
      k_c\in I\atop \bm k_v,\lambda\in\tau}\frac{1}{{\cal N}_v}\sum_{P_v}\delta(\varepsilon_{c\bm
  k_c}-\varepsilon_{v\bm k_v}\pm\hbar\Omega_{(\bm
  k_c-P_v\bm k_v)\lambda}\nonumber\\
&-&\hbar\omega)N_{(\bm k_c-P_v\bm
    k_v)\lambda\pm}\tilde{\cal A}^{abcd}_{I;c\bm k_cv(P_v\bm k_v)\lambda}\ .
\label{eq:X1}
\end{eqnarray}
Here $\tilde{\cal A}^{abcd}_{I;c\bm k_cv\bm
  k_v\lambda}=\sum_{P_{c,I}}{\cal A}^{abcd}_{c(P_{c,I}\bm
  k_c)v(P_{c,I}\bm k_v)\lambda}/{\cal N}_{c,I}$; $P_{c,I}$ are the
${\cal N}_{c,I}$  symmetry
operations in $C_{4v}$ that keep the $I^{th}$ valley unchanged, while
$P_v$ are the ${\cal N}_v$ symmetry operations in $O_h$; and
$\sum_{\lambda\in\tau}$ indicates summation over all modes in the
$\tau^{th}$ branch. Around the band edge, it is a good
approximation to take the mediated phonon energy $\hbar\Omega_{(\bm
  k_c-\bm k_v)\lambda}$ and the phonon number $N_{(\bm k_c-\bm
  k_v)\lambda}$ to be constant and equal to their band edge values $\hbar\Omega_{\bm
  k_c^0\lambda}=\hbar\Omega^0_{\tau}$ and $N_{\bm
  k_c^0;\lambda}=N_\tau^0$ , respectively. Then the injection rates are approximately
\begin{equation}
  {\cal A}^{abcd}_{I;cv\tau\pm} = \frac{2\pi}{\hbar}\sum_{\bm
      k_c\in I\atop \bm k_v}\delta(\varepsilon_{c\bm
  k_c}-\varepsilon_{v\bm
  k_v}\pm\hbar\Omega^0_{\tau}-\hbar\omega)N^0_{\tau\pm}\bar{\cal A}^{abcd}_{I;c\bm k_cv\bm k_v\tau}\ ,
\label{eq:X2}
\end{equation}
with 
\begin{eqnarray}
\bar{\cal A}^{abcd}_{I;c\bm k_cv\bm
  k_v\tau}&=&\frac{1}{{\cal N}_v}\sum_{P_v,\lambda\in\tau}\tilde{\cal A}^{abcd}_{I;c\bm k_cv(P_v\bm
  k_v)\lambda}\ .
\label{eq:symA}
\end{eqnarray}
Here the symmetrized expression $\bar{\cal A}^{abcd}_{I;c\bm k_cv\bm
  k_v\tau}$ in Eq.~(\ref{eq:X2})  avoids the ambiguity in calculating the band edge
values of ${\cal A}_{c\bm k_cv\bm
  k_v\lambda}^{abcd}$, which is induced by the degeneracy of the heavy and
light hole bands at the $\Gamma$ points. This can be
clearly shown by rewriting $W^{ab}_{\bar{c}\bm k_c\bar{v} \bm k_v\lambda}=\langle \bar{c}\bm
k_c|\hat{W}^{ab}_{c\bm k_cv\bm k_v\lambda} + \hat{W}^{ba}_{c\bm
  k_cv\bm k_v\lambda}|\bar{v}\bm k_v\rangle$ with the operator
\begin{eqnarray}
\hat{W}^{ab}_{c\bm k_cv\bm k_v\lambda} &\equiv& \frac{1}{2} (\frac{e}{\omega})^2
\bigg[H_{\lambda}^{ep}(\bm k_c-\bm
k_v)\frac{1}{H_e-\varepsilon_{v\bm
    k_v}-2\hbar\omega}\nonumber\\
&&\hspace{-1cm}\times\hat{v}^a\frac{1}{H_e-\varepsilon_{v\bm
    k_v}-\bar\omega}\hat{v}^b -\hat{v}^a\frac{1}{\varepsilon_{c\bm
      k_c}-H_e-\hbar\omega}\nonumber\\
&&\hspace{-1cm}\times H_{\lambda}^{ep}(\bm k_c-\bm
k_v)\frac{1}{H_e-\varepsilon_{v\bm k_v}-\hbar\omega}\hat{v}^b+\hat{v}^a\nonumber\\
&&\hspace{-1cm}\times\frac{1}{\varepsilon_{c\bm
    k_c}-H_e-\hbar\omega}\hat{v}^b\frac{1}{\varepsilon_{c\bm
    k_c}-H_e-2\hbar\omega} H_{\lambda}^{ep}(\bm k_c-\bm k_v)\bigg]\ ,\nonumber
\end{eqnarray}
keeping the intermediate states appearing in Eq.~(\ref{eq:W}) implicit.
Then similar to the corresponding results for one-photon indirect optical
transition\cite{Phys.Rev.B_83_165211_2011_Cheng}, we have 
\begin{equation}
  \bar{\cal A}^{abcd}_{I;c\bm k_c^0\text{HH}\bm
  k_v^0\tau} = \bar{\cal A}^{abcd}_{I;c\bm k_c^0\text{LH}\bm
  k_v^0\tau}=\frac{1}{2}\sum_{v^{\prime}=\text{LH},\text{HH}}\bar{\cal A}^{abcd}_{I;c\bm k_c^0v^{\prime}\bm
  k_v^0\tau}\ , 
\end{equation}
which is unambiguous for any choice of heavy and light hole state at
the valence band edge. We analyze the nonzero matrix elements of $\bar{\cal A}^{abcd}_{I;c\bm
  k_c^0v\bm k_v^0\tau}$ using the symmetries of the crystal.

For a given symmetry operation, the transformation of
$\hat{W}^{ab}$ is determined by $H^{ep}_{\lambda}$, $\hat{v}^a$, and
$\hat{v}^b$; a direct symmetry analysis for $W^{ab}_{\bar{c}\bm
  k_c^0\bar{v} \bm k_v^0\lambda}$ is possible with the electron state
$|\bar{c}\bm k_c^0\rangle$ and the hole state $|\bar{v}\bm
k_v^0\rangle$. However, because of very weak spin orbit coupling
in silicon, this process can be greatly simplified by dropping spin orbit
coupling terms in $H^{ep}_{\lambda}$, $\hat{v}^a$, and
$\hat{v}^b$, and thus in $\hat{W}^{ab}_{c\bm k_cv\bm k_v\lambda}$. Without spin-orbit coupling, the valence 
states at $\Gamma$ are chosen with the symmetry properties of $\{{\cal
  X}=yz, {\cal Y}=zx, {\cal Z}=xy\}$; the phonon states are chosen
with the symmetry $\{x,y\}$ for TA/TO branch, $\{z\}$ for LA branch, and
$\{x^2-y^2\}$ for LO branch; without losing generality, the conduction
band edge states are taken to lie in the $Z$ valley, which has the
symmetry of $\{z\}$. All matrix elements are listed in
Table~\ref{tab:nz-two}. In total there are fifteen nonzero quantities
for the band edge values. From the table, selection rules depend
strongly on phonon states.  

With spin orbit coupling, the valence bands are split into HH
($|\frac{3}{2},\pm\frac{3}{2}\rangle$), LH
($|\frac{3}{2},\pm\frac{1}{2}\rangle$) and SO
($|\frac{1}{2},\pm\frac{1}{2}\rangle$) bands, and the conduction bands
are two-fold spin degenerate bands $|z\uparrow\rangle$ and
$|z\downarrow\rangle$. The indirect optical matrix elements in these
states can be easily obtained by linear combination of the terms in
Table~\ref{tab:nz-two}, and the band edge transition probabilities can be identified
by $\bar{\cal A}^{abcd}_{Z;cv\tau}=\bar{\cal A}^{abcd}_{c\bm
  k_c^0v\bm k_v^0\lambda}$, with $\bm k_c^0$ being the band edge wave
vector in the $Z$ valley. Similar to the corresponding term for one-photon absorption\cite{Phys.Rev.B_83_165211_2011_Cheng}, $\bar{\cal
  A}^{abcd}_{Z;cv\tau}$ has the following properties: i) 
$\bar{\xi}^{abcd}_{Z;cv\tau}$ are the same for $v=$~HH, LH, and
SO; ii) $\sum_{v}\bar{\zeta}^{fabcd}_{Z;cv\tau}=0$ and
$\bar{\zeta}^{fabcd}_{Z;c\text{HH}\tau}=\bar{\zeta}^{fabcd}_{Z;c\text{LH}\tau}$. 
\begin{widetext}
\mbox{}
\begin{table}[htp]
  \centering
  \begin{tabular}[t]{|c|c|c|c|c|c|}
    \hline
    \hline
    &\multicolumn{2}{|c|}{TA/TO} & LA & LO \\
    \cline{2-5}
    \raisebox{1em}[0pt]{$W^{ab}_{c\bm k_c^0v\bm k_v^0\tau}$}&$x$&$y$&$z$&$x^2-y^2$\\
    \hline
    $|{\cal X}\rangle$ & $W_1^{(\prime)}M^{12}$ & $W_2^{(\prime)}M^{11} +
      W_3^{(\prime)}M^{22} + W_4^{(\prime)}M^{33}$& $W_6M^{23}$ &
    $W_8M^{13}$\\
    \hline
    $|{\cal Y}\rangle$ & $W_2^{(\prime)}M^{22} +
    W_3^{(\prime)}M^{11}+ W_4^{(\prime)}M^{33}$ &
    $W_1^{(\prime)}M^{12}$  &  $W_6M^{13}$ & $W_8M^{23}$\\
    \hline
    $|{\cal Z}\rangle$ & $W_5^{(\prime)}M^{23}$ &
    $W_5^{(\prime)}M^{13}$ & $W_7M^{12}$ & $W_9(M^{11} + M^{22}) + W_{10}M^{33}$\\
    \hline
    \hline
  \end{tabular}
  \caption{Band edge value of $W^{ab}_{c\bm k_c^0v\bm
      k_v^0\lambda}$. Here $M^{ij}$ is a matrix with matrix elements
    $[M^{ij}]_{kl}=(1-\delta_{ij})(\delta_{ik}\delta_{jl}+\delta_{il}\delta_{jk})
    + \delta_{ij}\delta_{ik}\delta_{il}$. There are totally fifteen
    parameters $\{W_i, i=1,\cdots,10\}$ for TA, LA, LO phonon branches
    and $\{W_i^{\prime},i=1,\cdots,5\}$ for the TO phonon branch.}
  \label{tab:nz-two}
\end{table}
\end{widetext}

\begin{table}[h]
  \centering
  \begin{tabular}[t]{|c|c|c|c|c|}
    \hline
    \hline
    &\multicolumn{3}{|c|}{$\tau$}\\
    \cline{2-4}
    \raisebox{1em}[0pt]{$\bar{\xi}_{Z;c\text{HH}\tau}^{abcd}$} &TA/TO & LA & LO \\
    \hline
    $\bar{\xi}_Z^{xxxx}$&$\frac{2}{3}(|W_3^{(\prime)}|^2+|W_2^{(\prime)}|^2)$&0&$\frac{2}{3}|W_9|^2$\\
    \hline
    $\bar{\xi}_Z^{xxyy}$&$\frac{4}{3}\text{Re}[W_3^{(\prime)}(W_2^{(\prime)})^{\ast}]$&0&$\frac{2}{3}|W_9|^2$\\
    \hline
    $\bar{\xi}_Z^{zzxx}$&$\frac{2}{3}W_4^{(\prime)}(W_3^{(\prime)}+W_2^{(\prime)})^{\ast}$&0&$\frac{2}{3}W_{10}W_9^{\ast}$\\
    \hline
    $\bar{\xi}_Z^{xyxy}$&$\frac{4}{3}|W_1^{(\prime)}|^2$&$\frac{2}{3}|W_7|^2$&0\\
    \hline
    $\bar{\xi}_Z^{xzxz}$&$\frac{2}{3}|W_5^{(\prime)}|^2$&$\frac{2}{3}|W_6|^2$&$\frac{2}{3}|W_8|^2$\\
    \hline
    $\bar{\xi}_Z^{zzzz}$&$\frac{4}{3}|W_4^{(\prime)}|^2$&0&$\frac{2}{3}|W_{10}|^2$\\
    \hline
  \end{tabular}
  \caption{Independent nonzero components of  $\bar{\xi}_{Z;c\text{HH}\tau}^{abcd}$. }
  \label{tab:nz-carrier}
\end{table}
\begin{table}[h]
  \centering
  \begin{tabular}[t]{|c|c|c|c|c|}
    \hline
    \hline
    &\multicolumn{3}{|c|}{$\tau$}\\
    \cline{2-4}
    \raisebox{1em}[0pt]{$\bar{\zeta}_{Z;c\text{HH}\tau}^{fabcd}$} &TA/TO & LA & LO \\
    \hline
    $\bar{\zeta}_Z^{zxyxx}$&$-\frac{i}{3}W_1^{(\prime)}(W_3^{(\prime)}-W_2^{(\prime)})^\ast$&0&0\\
    \hline
    $\bar{\zeta}_Z^{zyzxz}$&0&$-\frac{i}{3}|W_6|^2$&$\frac{i}{3}|W_8|^2$\\
    \hline
    $\bar{\zeta}_Z^{xyzyy}$&$\frac{i}{3}W_5^{(\prime)}(W_2^{(\prime)})^{\ast}$&0&$-\frac{i}{3}W_{8}W_9^{\ast}$\\
    \hline
    $\bar{\zeta}_Z^{xyzxx}$&$\frac{i}{3}W_5^{(\prime)}(W_3^{(\prime)})^{\ast}$&0&$-\frac{i}{3}W_{8}W_9^{\ast}$\\
    \hline
    $\bar{\zeta}_Z^{xxzxy}$&$\frac{i}{3}W_5^{(\prime)}(W_1^{(\prime)})^{\ast}$&$-\frac{i}{3}W_6W_7^{\ast}$&0\\
    \hline
    $\bar{\zeta}_Z^{xzzyz}$&$-\frac{i}{3}W_4^{(\prime)}(W_5^{(\prime)})^{\ast}$&0&$\frac{i}{3}W_{10}W_8^{\ast}$\\
    \hline
  \end{tabular}
  \caption{Independent nonzero components of $\bar{\zeta}_{Z;c\text{HH}\tau}^{fabcd}$.}
  \label{tab:nz-spin}
\end{table}

\begin{figure}[htp]
  \centering
  \includegraphics[width=8.5cm]{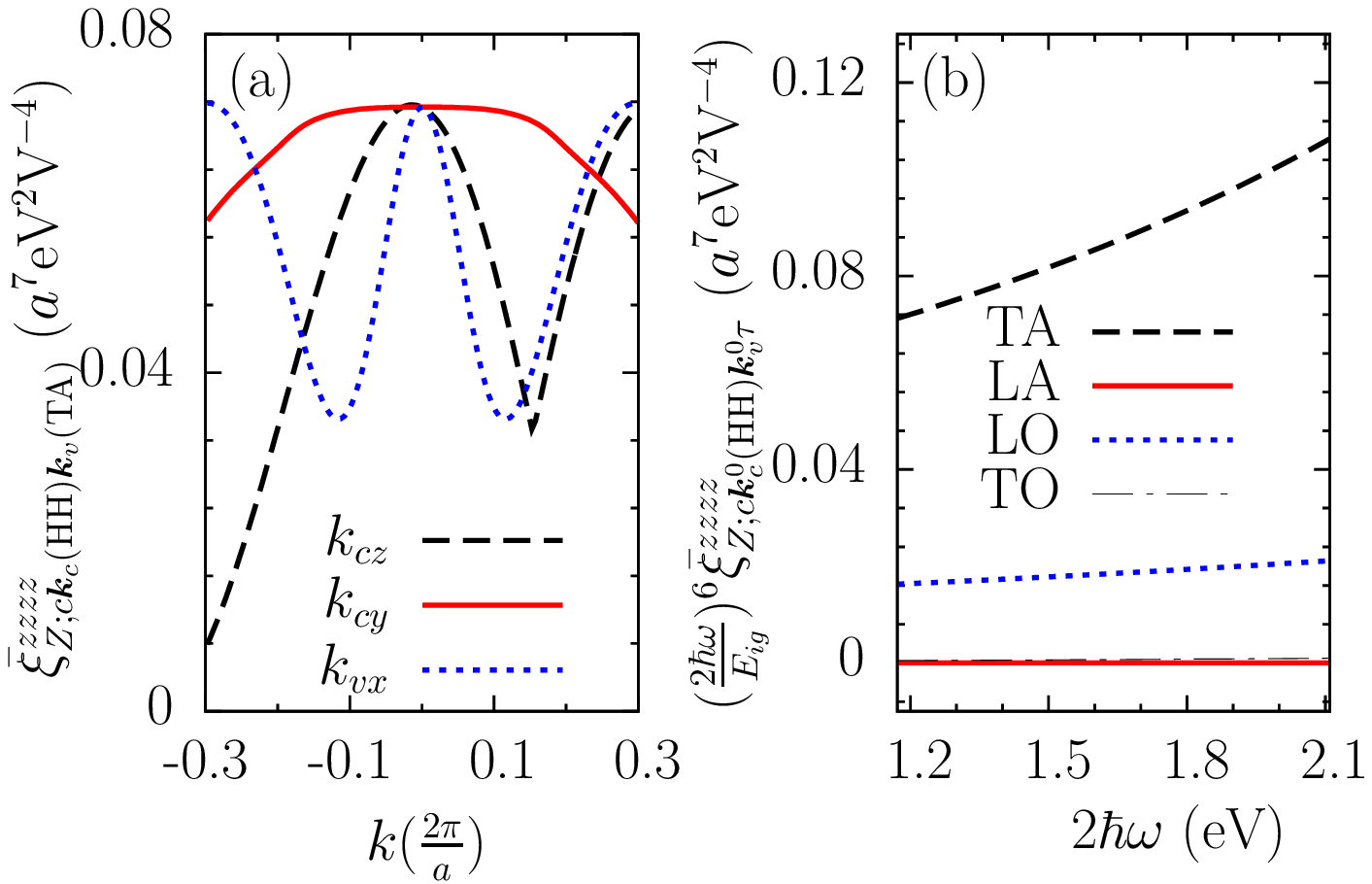}
  \caption{(color online). (a) Values of $\bar{\xi}_{Z;c\bm
      k_c(\text{HH})\bm k_v(\text{TA})}^{zzzz}$ at $2\hbar\omega=E_{ig}$ along different
    directions: (Dashed black curve) $(\bm k_c^0 + k\hat{\bm z},\bm k_v^0)$, 
    (Solid red curve) $(\bm k_c^0 + k\hat{\bm y},\bm k_v^0)$, and 
    (Dotted blue curve) $(\bm k_c^0,\bm k_v^0 + k\hat{\bm x})$. (b)
    photon energy dependence of $(2\hbar\omega)^6\bar{\xi}_{Z;c\bm
      k_c^0(\text{HH})\bm k_v^0\tau}^{zzzz}$ for different phonon
    branches. Here $a=5.431$\AA~is the lattice constant.}
\label{fig:bandedge_xik}
\end{figure}

We list $\bar{\cal A}^{abcd}_{Z;c\text{HH}\tau}$ in
Table~\ref{tab:nz-carrier} for carrier injection and 
Table~\ref{tab:nz-spin} for spin injection. 
Generally, these nonzero transition probabilities can be used in
Eq.~(\ref{eq:X2}) to approximate the $\bar{\cal A}^{abcd}_{I;c\bm
  k_cv\bm k_v\tau}$ around the band edge values, which results in a
simple formula 
\begin{equation}
  {\cal
    A}^{abcd}_{I;cv\tau\pm}\approx\frac{2\pi}{\hbar}J_{cv}(\hbar\omega)N^0_{\tau\pm}\bar{\cal
    A}^{abcd}_{I;cv\tau}\ ,
  \label{eq:aa}
\end{equation}
the analog of which is widely used in the qualitative analysis of
one-photon  direct and indirect injection even for injection away from the band edge. Here $J_{cv}(\hbar\omega)$
is the joint density of states, $J_{cv}(\hbar\omega)=\sum_{\bm
  k_c\in I; \bm k_v}\delta(\varepsilon_{c\bm k_c}-\varepsilon_{v\bm
  k_v}\pm\hbar\Omega^0_{\tau}-\hbar\omega)$. In
Fig.~\ref{fig:bandedge_xik}(a), we give the local properties of
$\bar{\xi}_{Z;c\bm k_c(\text{HH})\bm k_v\tau}^{xxxx}$ around band
edges $(\bm k_c^0,\bm k_v^0)$; its rapid variation away from the band
edge shows that the simple formula (\ref{eq:aa}) may fail.
Garcia and Kalyanaraman\cite{J.Phys.B_39_2737_2006_Garcia} found
that the corresponding formula for two photon absorption should be replaced by
\begin{equation}
  \beta=\sum_{n\lambda\pm}C_{n\lambda\pm}F_{n}\left(\frac{2\hbar\omega}{E_{ig}},
    \frac{\pm\hbar\Omega_{\lambda}^0}{E_{ig}}\right)\ .
\label{eq:fitformula}
\end{equation}
Here the two-photon absorption coefficient $\beta$ is related to our
calculated quantity $\xi^{xxxx}$ by
$\beta=2\hbar\omega\xi^{xxxx}/(2n_Rc\epsilon_0)^2$, $n_R$
is the refractive index, $c$ is the speed of light, $\epsilon_0$
is the vacuum permittivity, $F_{n}(x,y) = (x-y-1)^{2+n}/x^5$, and
$\Omega_{\lambda}^0=\Omega_{\bm   k_c^0\lambda}$ is the frequency of
phonons mediated in the band edge transitions. According to the parity
difference between the band edge hole and electron states, the transitions are
divided into allowed-allowed ($a-a$), allowed-forbidden ($a-f$), and
forbidden-forbidden ($f-f$) processes, which correspond to the $n=0$,
1, and 2 terms in Eq.~(\ref{eq:fitformula}), respectively; such a 
classification is based on whether the band edge values of the matrix
elements of $v^a$ and $v^b$ in Eq.~(\ref{eq:W}) are zero (forbidden) or
nonzero (allowed) for different parity of the intermediate
states. In deriving Eq.~(\ref{eq:fitformula}), the dependence $\xi^{zzzz}_{c\bm k_cv\bm
  k_v\lambda}\propto (\hbar\omega)^{-6}$ must be
used. Dinu\cite{IEEEJ.QuantumElectron._39_1398_2003_Dinu} argued
instead that $\xi^{zzzz}_{c\bm k_cv\bm k_v\lambda}\propto (\hbar\omega)^{-5}$ for
some processes. Here we can numerically study this dependence,  and  the
result is plotted in Fig.~\ref{fig:bandedge_xik}(b); it shows that 
$\omega^{-5}$ and $\omega^{-6}$ dependences are both important, at least for the TA phonon branch. 

The simple formula (\ref{eq:aa}) only corresponds to the $a-a$
process. In obtaining results for the other two processes, the first and second
derivatives with respect to $\bm k_c$ and $\bm k_v$ of $W^{ab}_{I;c\bm k_cv\bm
  k_v\lambda}$ are necessary. This results in a more complicated symmetry
analysis that we do not consider here.

\subsubsection{Injection for $\sigma^-$ light propagating along
  $\langle 001\rangle$ and $\langle 111\rangle$}

For $\sigma^-$ light propagating along the directions $\langle 001\rangle$ and
$\langle 111 \rangle$, the electric field $\bm E^{\langle \hat{\bm
    k}\rangle}$ can
be written respectively as
\begin{eqnarray}
\bm E^{\langle 001 \rangle}_{\omega}&=&\frac{\hat{\bm
    x}-i\hat{\bm y}}{\sqrt{2}}E_0\ ,\nonumber\\
\bm E^{\langle 111 \rangle}_{\omega}&=&\frac{2i\hat{\bm x} +
(\sqrt{3}-i)\hat{\bm y}
-(\sqrt{3}+i)\hat{\bm z}}{2\sqrt{3}}E_0\ ,
\end{eqnarray}
where $\langle \hat{\bm k}\rangle$ denotes $\langle001\rangle$ or $\langle111\rangle$.
The injection rates of carriers and spins then are
\begin{eqnarray}
  \dot{n}_{I;cv\lambda\pm}&=&\xi^{\langle\hat{\bm k}\rangle}_{I;cv\lambda\pm}|E_0|^4\ ,\nonumber\\
  \dot{S}^{f}_{I;cv\lambda\pm}&=&\zeta^{f;\langle\hat{\bm k}\rangle}_{I;cv\lambda\pm}|E_0|^4\ .
\end{eqnarray}
Here $\xi_I^{\langle\hat{\bm k}\rangle}$  and $\zeta_I^{f;\langle\hat{\bm k}\rangle}$ are the injection
coefficients of carriers and spins, respectively, in the $I^{th}$ valley.
They can be expressed by the nonzero components of
$\xi^{abcd}_{Z;cv\lambda\pm}$ and $\zeta^{fabcd}_{Z;cv\lambda\pm}$
defined in Eqs.~(\ref{eq:nz-xi}) and (\ref{eq:nz-zeta}). The resulting
expressions are listed in Table~\ref{tab:rateforel}. The DSP is
defined as $\mathtt{DSP}^{f}=\dot{S}^{f}/(\hbar\dot{n}/2)$. For $\langle
001\rangle$ light, the injected spin in each valley
 and  the total spin are all parallel to the light propagation direction,
 {\it i.e.}, the $\langle001\rangle$ direction. The carrier and
spin injection rates show valley anisotropy between the $Z$ valley and
the $X/Y$ valleys. For $\langle 111 \rangle$
light, the injected carriers are the same for every
valley, and the total spin polarization is still along the direction
of the electric field, {\it i.e.}, the $\langle 111\rangle$ direction. But
the injected spins in each valley have different spin polarization:
The two transverse directions in each valley have the same injection
rates, which are different from the longitudinal direction of the valley.

\begin{widetext}

\begin{table}[htp]
    \begin{tabular}[t]{|c|c|c|c|c|c|}
      \hline
      $\langle\hat{\bm
        k}\rangle$&$I$&$\xi_I^{\langle\hat{\bm k}\rangle}$&$\zeta_I^{x;\langle\hat{\bm
          k}\rangle}$&$\zeta_I^{y;\langle\hat{\bm
          k}\rangle}$&$\zeta_I^{z;\langle\hat{\bm k}\rangle}$\\
      \hline
      &$X$&$\frac{1}{4}(\xi^{zzzz}_Z+\xi^{xxxx}_Z)-\frac{1}{2}\xi^{zzxx}_Z
      +
      \xi^{zxzx}_Z$&0&0&$\text{Im}[\zeta^{xzzyz}_Z+\zeta^{xyzyy}_Z]$\\
      \cline{2-6}
      &Y&$\xi^{\langle 001\rangle}_X$&0&0&$\zeta^{z;\langle 001\rangle}_Z$\\
      \cline{2-6}
      \raisebox{1em}[0pt]{$\langle 001\rangle$}&$Z$&$\frac{1}{2}\xi^{xxxx}_Z-\frac{1}{2}\xi^{xxyy}_Z+\xi^{xyxy}_Z$&0&0&$2\text{Im}[\zeta^{zxyxx}_Z]$\\
      \cline{2-6}
      &Total&$\frac{1}{2}\xi^{xxxx}-\frac{1}{2}\xi^{xxyy}+\xi^{xyxy}$&0&0&$2\text{Im}[\zeta^{zxyxx}]$\\
      \hline
      &$X$&$\xi_Z^{\langle 111\rangle}$&$\zeta_Z^{z;\langle 111\rangle}$&$\zeta_Z^{x;\langle 111\rangle}$&$\zeta_Z^{x;\langle 111\rangle}$\\
      \cline{2-6}
      $\langle 111\rangle$&Y&$\xi_Z^{\langle 111\rangle}$&
      $\zeta_Z^{x;\langle 111\rangle}$&$\zeta_Z^{z;\langle 111\rangle}$&$\zeta_Z^{x;\langle 111\rangle}$\\
      \cline{2-6}
      &$Z$&$\frac{1}{9}\left[2\xi^{xxxx}_Z+\xi^{zzzz}_Z-\xi^{xxyy}_Z-2\xi^{xxzz}_Z
        \atop+
        4\left(\xi^{xyxy}_Z+2\xi^{xzxz}_Z\right)\right]$&$
      \frac{2}{3\sqrt{3}}\text{Im}[\zeta^{xyzyy}_Z+2\zeta^{xxzxy}_Z+\zeta^{xzzyz}_Z]$&$\zeta_Z^{x;\langle
        111\rangle}$&$\frac{4}{3\sqrt{3}}\text{Im}[\zeta^{zxyxx}_Z+\zeta^{zyzxz}_Z]$\\
      \cline{2-6}
      &Total&$\frac{1}{3}\left(\xi^{xxxx}-\xi^{xxyy}+4\xi^{xyxy}\right)$&$\frac{4}{3\sqrt{3}}\text{Im}[\zeta^{zxyxx}+\zeta^{xxzxy}]$&$\zeta^{x;\langle
        111\rangle}$&$\zeta^{x;\langle 111 \rangle}$\\
      \hline
    \end{tabular}
    \caption{The carrier indirect two-photon injection coefficients
      $\xi_I^{\langle\hat{\bm k}\rangle}$ and the spin indirect two-photon injection
      coefficients $\zeta^{a;\langle\hat{\bm k}\rangle}_I$ in the $I^{th}$ valley for
      $\sigma^-$ light propagating along directions $\langle \hat{\bm k}\rangle=\langle 001 \rangle $ and $\langle 111 \rangle$.}
    \label{tab:rateforel}
\end{table}
\end{widetext}

\subsubsection{Carrier injection under $\sigma^-$ light propagating along
  $\langle 001\rangle$ and $\langle 111\rangle$}
\begin{figure}[htp]
  \centering
  \includegraphics[width=8.5cm]{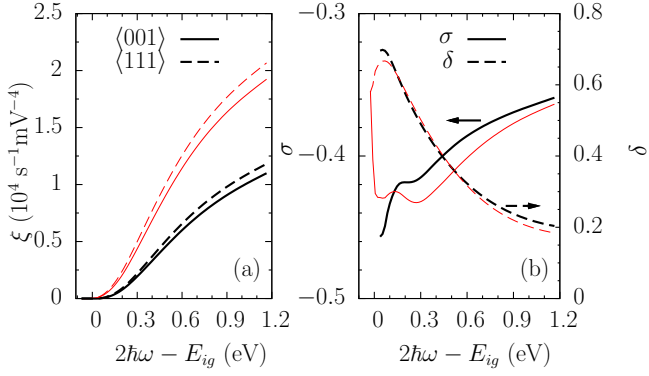}
  \caption{(Color online) (a) Spectra of total carrier injection rates $\xi$
    at 4~K and 300~K for $\sigma^-$ light propagating along the $\langle
    001\rangle$ and $\langle 111 \rangle$ directions. (b) Anisotropy
    $\sigma$ and linear-circular dichroism $\delta$ at 4~K (thick black
    curves) and 300~K (thin red curves).}
\label{fig:carriertotal}
\end{figure}

We plot photon energy dependence of the total carrier injection
coefficients for  $\langle001\rangle$ and $\langle 111\rangle$ light at 4~K and 300~K
in Fig.~\ref{fig:carriertotal} (a). As we found earlier in a
  preliminary study\cite{Appl.Phys.Lett._98_131101_2011_Cheng}, the injection
coefficients increase rapidly with increasing temperature. The injection
for $\langle 111\rangle$ light is larger than that for
$\langle001\rangle$ light, demonstrating the anisotropy of the 
injection on light propagating direction. In agreement with Hutchings and
Wherrett's notation\cite{Phys.Rev.B_49_2418_1994_Hutchings} for direct gap two-photon injection, this anisotropy
can be characterized by two parameters, the anisotropy $\sigma$ and
the linear-circular dichroism $\delta$, which are given as
\begin{eqnarray}
  \sigma &=& \frac{\xi^{xxxx}-2\xi^{xyxy}-\xi^{xxyy}}{\xi^{xxxx}}\
  ,\nonumber\\
  \delta &=& \frac{\xi^{xxxx}-2\xi^{xyxy}+\xi^{xxyy}}{2\xi^{xxxx}}\ .
  \label{eq:dichroism}
\end{eqnarray}
In the isotropic limit, $\sigma=0$ and
$\delta=\xi^{xxyy}/\xi^{xxxx}$\cite{Opt.Mater._3_53_1994_Hutchings}. We plot $\sigma$ and $\delta$ in
Fig.~\ref{fig:carriertotal} (b). Note that the anisotropy shows a much
stronger temperature dependence than the linear-circular dichroism. In contrast to $\sigma$ in direct
gap two-photon injection, which clearly shows the onset of the
transition from the spin split-off band to the conduction band by the
presence of a cusp, it is hard to identity the contribution from spin split-off band in
indirect gap injection. This is because the energy dependence at the
onset of indirect absorption is $\propto(\hbar\omega-E_{ig})^2$, given
by the $a-a$ process, instead of $\propto(\hbar\omega-E_{g})^{1/2}$
for direct absorption. 

\begin{figure}[htp]
  \centering
  \includegraphics[width=6cm]{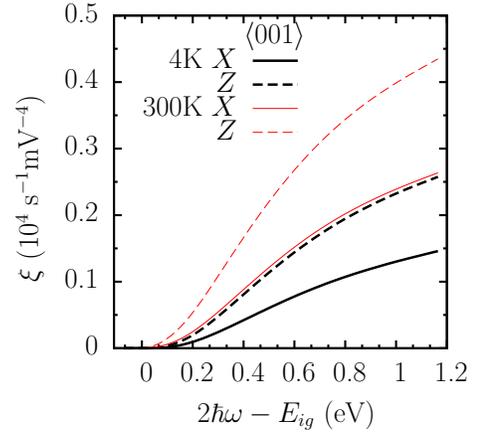}
  \caption{(Color online) Spectra of
    carrier injection rates $\xi_{I}^{\langle 001\rangle}$ in the $I=Z$ and $X$
    valley at 4 (thick black curves) and 300~K (thin red curves).}
\label{fig:carriervalley001}
\end{figure}
Now we turn to the carrier injection into each valley. For $\langle
111\rangle$ light, all valleys are equivalent, and the injection
coefficient in each valley is $1/6$ of the total. There is no valley
anisotropy in this case. For $\langle 001\rangle$ light, the valleys
can be divided into two sets: $\{Z, \bar{Z}\}$ 
and $\{X,\bar{X},Y,\bar{Y}\}$, and the injection is the same for all
valleys within each set. We plot the spectra of injection rates
$\xi_I^{\langle 001\rangle}$ in the $I=Z$ and $X$ valleys at 4~K and
300~K in Fig.~\ref{fig:carriervalley001}. The spectrum in each valley has a
 shape similar to the total, and the injection in the $Z$ valley is
larger than that in the $X$ valley. The valley anisotropy arises
because of the anisotropic effective electron mass in the conduction
bands, which leads to different matrix elements appearing in
(\ref{eq:W}) for the different Cartesian components of velocity. For
the $Z$ valley, the effective mass along the $z$ direction is
heavier than that along the $x/y$ directions\cite{landolt_Si}, which
results in a smaller $z$ component of the interband velocity matrix
elements\cite{effectivemass}. From Table~\ref{tab:rateforel}, we 
find that the $z$-components of the electron and hole velocity
only appears in the injection rates in the $X/Y$ valleys,
and results in their smaller injection rates.

\begin{figure}[htp]
  \centering
  \includegraphics[width=7cm]{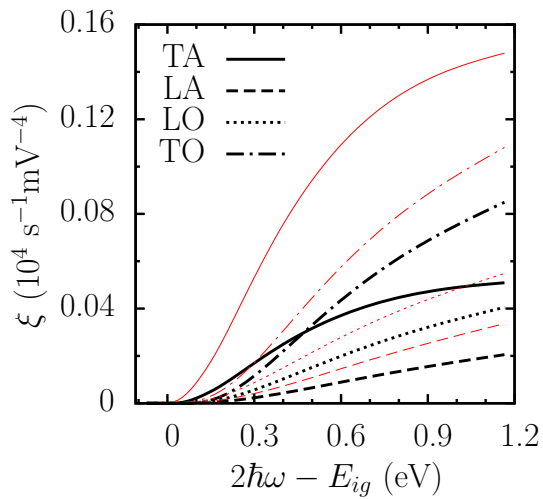}
  \caption{(Color online) Phonon resolved spectra of carrier
    injection rates $\xi_{X;\tau}^{\langle 111\rangle}$  at 4~K (thick
    black curves) and 300~K (thin red curves).} 
\label{fig:carrierphonon111}
\end{figure}
Figure \ref{fig:carrierphonon111} gives the phonon-resolved spectra in
the $X$ valley for  $\langle111\rangle$ light. Similar to our previous
results\cite{Appl.Phys.Lett._98_131101_2011_Cheng}, we find here that the LA
phonon-assisted process gives the smallest contribution, while the TA 
and TO phonon-assisted processes dominate: At low temperature, the TA 
phonon-assisted process dominates at low photon energy, and the TO
phonon-assisted process dominates at high photon energy; with
temperature increasing, the TA phonon-assisted process becomes more and more
important due to the small TA phonon energy, and dominates for
photon energy less than $E_{ig}$ at 300~K. The phonon-resolved
injection rates in each valley for $\langle001\rangle$ light show similar behavior.

\subsubsection{Spin injection under $\sigma^-$ light propagating along
  $\langle 001\rangle$ and $\langle 111\rangle$}
\begin{figure}[htp]
  \centering
  \includegraphics[width=8cm]{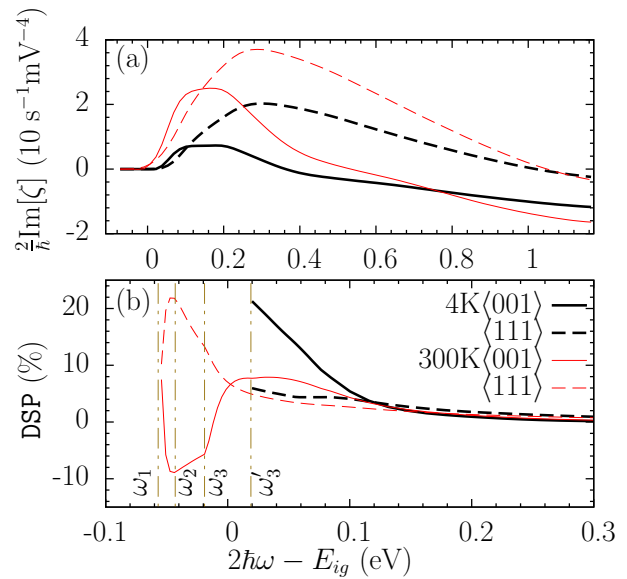}
  \caption{(Color online) Spectra of (a) spin injection rates
    $\zeta^{\langle\hat{\bm k}\rangle}$ and (b) $\text{DSP}^{\langle\hat{\bm k}\rangle}$  at 4~K (black
    curves) and 300~K (red curves).  (Solid curves) $\bm
    E\parallel\langle 001\rangle$, (dashed curves) $\bm
    E\parallel\langle 111\rangle$. The spin polarization direction is
    parallel to $\bm E$. The labeled energies are
    $2\hbar\omega_{1-3}-E_{ig}=-\Omega_{\text{TO}}^0$,
    $-\Omega_{\text{LA}}$, $-\Omega_{\text{TA}}$, and $2\hbar\omega_{3}^{\prime}-E_{ig}=\Omega_{\text{TA}}$.} 
  \label{fig:spintotal}
\end{figure}
In Fig.~\ref{fig:spintotal} we show the spectra of the spin injection
rates $\zeta^{\langle\hat{\bm k}\rangle}$ and the $\text{DSP}^{f}$ for $\sigma^-$
light propagating along $\langle 001\rangle$ and $\langle 111\rangle$
directions at 4~K and 300~K. The total spin polarizations are all parallel to
light propagation direction. When photon energy is higher than
the injection edge, which is $E_{ig}+\hbar\Omega_{\text{TA}}^0$ at 4~K or
$E_{ig}-\hbar\Omega_{\text{TO}}^0$ at 300~K, the spin injection rates first
increase with photon energy from zero to maximum values, then
decrease, and then change direction at high photon energies. 
This is different than the behavior of indirect one-photon spin injection, in which the  
spin injection rates always increase with photon energy. The
difference can be attributed to the complicated transition amplitude $W$ in
Eq.~(\ref{eq:W}). The fine structure of the injection rates around the
band edge are clearly shown by the DSP spectra in Fig.~\ref{fig:spintotal}(b). 
The DSP depends strongly on the laser propagation
direction and the temperature. For $\langle 001\rangle$ light, the maximum
DSP can reach about $20\%$ at 4~K and $-10\%$
at 300~K; for $\langle 111\rangle$ light, the
maximum DSP is only $6\%$ at 4~K but $20\%$ at 300~K. Around the
injection edges, the DSP show more detailed structures at 300~K than
4~K.  In Fig.~\ref{fig:spintotal}(b), we label the injection edge for phonon
branches by dotted vertical lines: $\hbar\omega_1$, $\hbar\omega_2$,
and $\hbar\omega_3$ for the TO, LO, and TA phonon absorption process,
respectively; $\hbar\omega_3^{\prime}$ identifies the TA phonon emission
process. As in the corresponding results for indirect one-photon spin injection, the fine
structures arising here come from the contributions of different phonon
branches. 
\begin{figure}[htp]
  \centering
  \includegraphics[width=8cm]{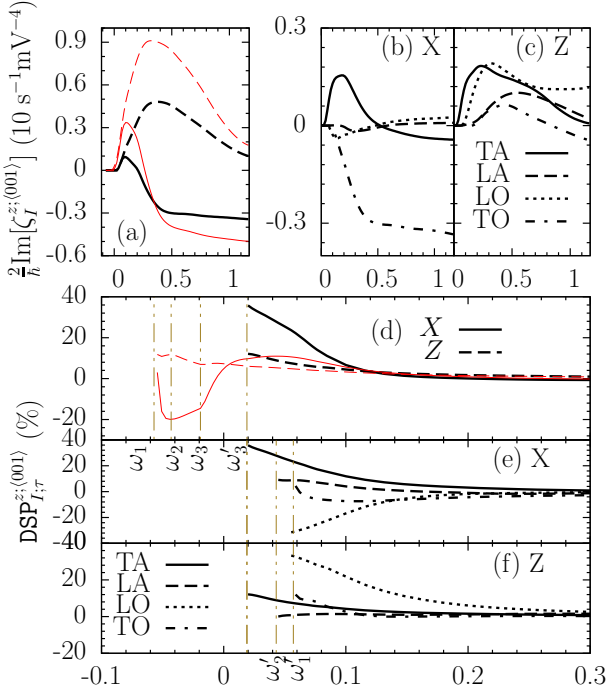}
  \caption{(Color online) Spectra of (a) spin injection rates
    $\zeta_I^{\langle 001\rangle}$ in the $I=X$ valley (solid curves)
    and the $I=Z$ valley (dashed curve) at 4~K (thick black curves) and
    300~K (thin red curves), (b)-(c) phonon branch resolved spin
    injection rates in the $X$ and $Z$ valleys at 4~K, and (d-f) The
    corresponding DSP of (a-c).The labeled energies are
    $2\hbar\omega_{1-2}^{\prime}=\Omega_{\text{TO}}^0$ and $\Omega_{\text{LA}}^0$, respectively.}
\label{fig:spin001}
\end{figure}

To better understand these fine structures, we plot the spin injection in each valley
and the contribution from each phonon branches for $\langle001\rangle$ light in Fig.~\ref{fig:spin001}. Figure~\ref{fig:spin001}(a) gives the spin injection
rates in the $X$ and $Z$ valleys at 4~K and 300~K, in which the valley
anisotropy is prominent. Again, due to the anisotropy in electron 
velocity, the injection rate in the $Z$ valley is much larger
than that in the $X$ valley. Figure~\ref{fig:spin001}~(d)
gives the detailed structure of DSP around the band edge. The maximum DSP is
about $36\%$ at 4~K and $-20\%$ at 300~K in the $X$ valley, and about
$10\%$ in the $Z$ valley for both temperatures. However, the spin injection rates are
very close in these two valleys near the injection edge in Fig.~\ref{fig:spintotal}(a), so the
difference between these maximum values can only come from the difference of
the carrier injection rates, which are much smaller in the $X$
valley than that in the $Z$ valley (see
Fig.~\ref{fig:carriervalley001}). 

The phonon-resolved spin injection rates in the $X$ and $Z$ valleys are plotted in
Figs.~\ref{fig:spin001}~(b-c) at 4~K . In the $X$ valley, the TA phonon
branch dominates at low photon energy, and the TO phonon branch
dominates at high photon energy. In the $Z$ valley, the TA and LO
phonon branches have similar contributions and dominate at low photon
energy. Near the band edge, the spins injected from the TO and TA
phonon-assisted processes have opposite spin polarization direction in the
$X$ valley, but  same in the $Z$
valley. Figures~\ref{fig:spin001}~(e-f)
give the corresponding DSP. Almost all processes contribute nonzero
DSP. In the $Z$ valley, the spin injection rates are given by
$\text{Im}[\zeta_{Z;\tau}^{zxyxx}]$ as shown in
Table~\ref{tab:rateforel}. At the band edges, the carrier and spin injection amplitude of the LA and
LO phonon-assisted processes are all zero, which means these $a-a$
processes inject no carriers. As one moves away from the band edges,
carriers and spins can be injected by $a-f$ and $f-f$ processes, which
results in a nonzero DSP. 
In the $X$ valley, the spin injection rates are given by
$\text{Im}[\zeta_{Z;\tau}^{xzzyz}+\zeta_{Z;\tau}^{xyzyy}]$. From the
results in Table~\ref{tab:nz-spin}, we see that the LA phonon-assisted
process gives zero spin injection amplitude at band edge, but its DSP
is not zero because the $a-f$ and $f-f$ processes dominate over the $a-a$
process. Similar results also exist in two-photon direct injection\cite{Phys.Rev.B_71_035209_2005_Bhat}. 

In Figs.~\ref{fig:spin001}~(e-f), we plot only the DSP for the phonon emission processes, the corresponding injection
edges are given by $\hbar\omega_i^{\prime}$. From the calculation of the indirect
one-photon injection, we know that the DSP induced by the phonon 
absorption process and the phonon emission process have a similar
shape, but the injection edge shifts from $\hbar\omega_i^{\prime}$  to
$\hbar\omega_i$. At 4~K, the injection edge is dominated by the TA
phonon emission process (which begins at
$\hbar\omega_3^{\prime}$), and it is dominated by the TO/LO phonon
absorption process (begins at $\hbar\omega_1$) at 300~K, then follows
by the LA phonon absorption process at $\hbar\omega_2$ and the TA phonon at
$\hbar\omega_3$. Therefore, the co-action of the TO and LO phonon
absorption processes gives the negative DSP, and results in the sharp increase between the photon
energies $\hbar\omega_1$ and $\hbar\omega_2$ in
Fig.~\ref{fig:spin001}~(c), then the LA/TA phonon absorption processes
give positive DSP, so the total DSP decreases sharply after
$\hbar\omega_2$. 

\begin{figure}[htp]
  \centering
  \includegraphics[width=8cm]{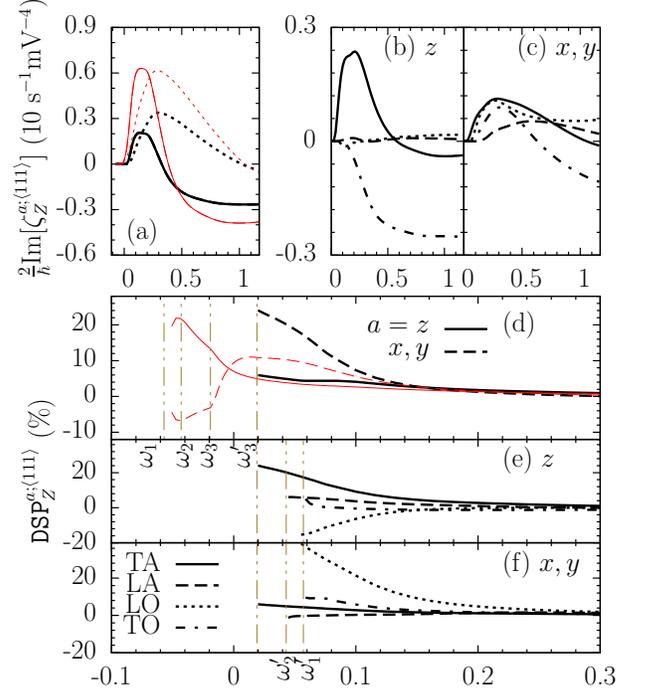}
  \caption{(Color online) Spectra of (a) spin injection rates
    $\zeta_Z^{a;\langle 111\rangle}$ and (d) $\text{DSP}_Z^{a;\langle
      111\rangle}$ for the $a=z$ (solid curves) and $x,y$ (the
     dashed curves) spin components at 4~K (thick black
    curves) and 300~K (thin red curves). Phonon resolved (b-c) spin injection rates
    $\zeta_{Z;\tau}^{a;\langle 111\rangle}$ and (e-f) $\text{DSP}_{Z;\tau}^{a;\langle
      111\rangle}$ for the $z$ (black thick curves) and $x,y$ (blue
    thin curves)
    spin components at 4~K.}
\label{fig:spin111}
\end{figure}

Figure~\ref{fig:spin111} gives the details of the spin injection for
the $\sigma^-$ light propagating along $\langle111\rangle$
direction. The analysis is similar to the $\langle001\rangle$ case.

\section{\label{sec:coherentcontrol}Two-color charge and spin current
  injection}
\begin{figure}[htp]
  \centering
  \includegraphics[width=8cm]{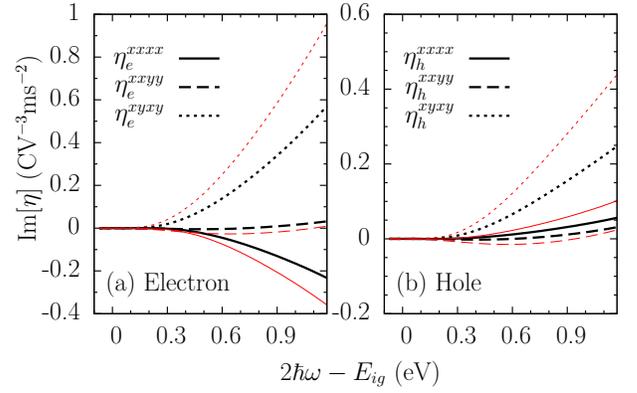}
  \caption{(Color online) Spectra of $\text{Im}[\eta^{fabc}]$
    for electron (a) and hole (b) at 4~K (thick black curves) and
    300~K (thin red curves).}
\label{fig:chargecurrent}
\end{figure}

\begin{figure}[htp]
  \centering
  \includegraphics[width=8cm]{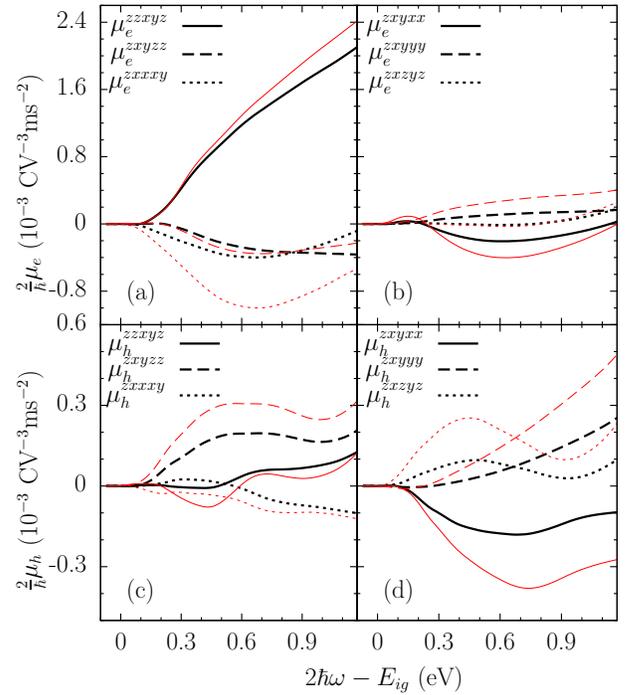}
  \caption{(Color online) Spectra of $\mu_{e}^{dfabc}$ for electron
    (a) and hole (b) at 4~K (thick black curves) and  300~K (thin red
    curves)} 
\label{fig:spincurrent}
\end{figure}

Now we study the motions of optically injected carriers and spins. 
While a single color light source can inject net current into a  particular
valley\cite{JETPLett._81_231_2005_Tarasenko,Phys.Rev.B_83_121312_2011_Karch},
due to the $O_h$ symmetry there is no net charge or spin current
injection from either either one-photon or two-photon
absorption of a single color light source. We
calculate $1+2$ injection effects here, and only consider the total
charge and spin current induced.
For a two-color optical field ${\bm E}(t) = {\bm
  E}_\omega e^{-i\omega t} + {\bm E}_{2\omega}e^{-i2\omega t}+ c.c$,
the carrier density injection rate is 
\begin{eqnarray}
  \dot{n} &=& \xi^{ab}E_{2\omega}^a(E_{2\omega}^b)^{\ast} +
  \xi^{abcd}E_{\omega}^aE_{\omega}^b (E_\omega^cE_\omega^d)^{\ast}\ ,
\label{eq:totaln}
\end{eqnarray}
where $\xi_i^{ab}$ are one-photon indirect injection
coefficients\cite{Phys.Rev.B_83_165211_2011_Cheng} and $\xi^{abcd}$
are the two-photon injection coefficients studied in the previous sections. The interference between $\omega$ and $2\omega$ beams
injects charge and spin currents with injection rates
\begin{eqnarray} 
\dot{J}^{d}_{e(h)} &=& \eta_{e(h)}^{dabc}E^a_{2\omega}(E^b_\omega
E^c_\omega)^{\ast} + c.c.\ ,\nonumber\\
\dot{K}^{fd}_{e(h)} &=&\mu_{e(h)}^{dfabc}E^a_{2\omega}(E^b_\omega
E^c_\omega)^{\ast} + c.c.\ .
\end{eqnarray}
The injection coefficients $\eta_{e(h)}^{dabc}$ and
$\mu^{dfabc}_{e(h)}$ are written in the form ${\cal
  B}^{abc}=\sum_{\lambda\pm}{\cal B}^{abc}_{\lambda\pm}$ with 
{\allowdisplaybreaks
\begin{eqnarray}
  {\cal B}^{abc}_{\lambda\pm} &=& \frac{2\pi}{\hbar}\sum_{
    c\bm k_c, v\bm k_v}\delta(\varepsilon_{c\bm
  k_c}-\varepsilon_{v\bm k_v}\pm\hbar\Omega_{(\bm
  k_c-\bm k_v)\lambda}-2\hbar\omega) \nonumber\\
  &\times&N_{(\bm k_c-\bm k_v)\lambda\pm} {\cal B}^{abc}_{c\bm k_cv\bm k_v\lambda}\ ,\\
  {\cal B}^{abc}_{c\bm k_cv\bm k_v\lambda} &=& i\sum_{\sigma_c\sigma_c^{\prime};\sigma_v\sigma_v^{\prime}}\langle
  \bar{c}^{\prime}\bm k_c|\langle \bar{v}^{\prime}\bm k_v|\hat{\cal B}|\bar{v}\bm
  k_v\rangle|\bar{c}\bm  k_c\rangle \nonumber\\ 
  &\times&  T^{a}_{\bar{c}\bm k_c\bar{v}\bm 
    k_v\lambda}(2\omega)[W^{bc}_{\bar{c}^{\prime}\bm k_c\bar{v}^{\prime}\bm
    k_v\lambda}(\omega)]^{\ast}\ .
\end{eqnarray}
}
Here $T^a_{\bar{c}\bm k_c\bar{v}\bm k_v\lambda}$ is the one-photon indirect
optical transition amplitude
\cite{Phys.Rev.B_83_165211_2011_Cheng}. By taking $\hat{\cal B}$ as
$\hat{J}_{e}^{d}=-e\hat{v}_e^{d}$, 
$\hat{J}_h^{d}=e\hat{v}_h^d$,
$\hat{K}_{e}^{fd}=-\frac{e}{\hbar}(\hat{v_e}^d\hat{S}_e^{f}+\hat{S}_e^f\hat{v}_e^d)$,
and $\hat{K}_{h}^{fd}=\frac{e}{\hbar}(\hat{v_h}^d\hat{S}_h^{f}+\hat{S}_h^f\hat{v}_h^d)$,
we obtain the injection rates for electron and hole charge and spin
currents, with $\eta^{dabc} = \eta^{dacb}$ and $\mu^{dfabc} =
\mu^{dfacb}$. For diamond structure crystals, the nonzero components are
\begin{eqnarray}
\eta^{xxxx} &\ ,&\nonumber\\
\eta^{xxyy} &=&\eta^{xxzz}\ ,\nonumber\\
\eta^{xyxy} &=&\eta^{xzxz}\ .
\end{eqnarray}
and
\begin{eqnarray}
\mu^{zxxxy} &=& -\mu^{zyyyx}\ ,\nonumber\\
\mu^{zxyxx} &=& -\mu^{zyxyy}\ ,\nonumber\\
\mu^{zxyyy} &=& -\mu^{zyxxx}\ ,\nonumber\\
\mu^{zxyzz} &=& -\mu^{zyxzz}\ ,\nonumber\\
\mu^{zxzyz} &=& -\mu^{zyzxz}\ ,\nonumber\\
\mu^{zzxyz} &=& -\mu^{zzyxz}\ .
\end{eqnarray}
All other nonzero components can be obtained by cyclic permutations of
the Cartesian indices. The phonon-resolved tensors $\eta^{dabc}_{e(h);\lambda\pm}$ and
$\mu^{dfabc}_{e(h);\lambda\pm}$ share the same symmetry
properties as the total injection tensor $\eta^{dabc}_{e(h)}$ and
$\mu^{dfabc}_{e(h)}$, respectively. Using time-reversal symmetry, in
the independent particle approximation we adopt here all $\eta^{dabc}$ are pure imaginary numbers, and all $\mu^{dfabc}$ are real numbers. We show the calculated spectra of each 
component of the charge current in Fig.~\ref{fig:chargecurrent}, and of
the spin current in Fig.~\ref{fig:spincurrent}. The current injection coefficients $\eta_{e(h)}^{dabc}$ and
$\mu_{e(h)}^{dfabc}$ have the same symmetry properties as that of the
two-color direct current injection across the direct gap of germanium
\cite{Phys.Rev.B_81_155215_2010_Rioux}. 

From the calculation, both the charge and the spin currents for
injected electrons are larger than that for injected holes. One
contributing factor is that the electron moves faster than
the hole due to the smaller effective mass. But for the spin current,
another factor is that the average spin expectation value
over the HH and LH band is smaller than that in the conduction bands.

We consider the indirect current injection coefficients under the two-color laser beams propagating
along the $z$ direction with the electric field components taken as $\bm E_\omega =
E_\omega e^{i\phi_{\omega}}\hat{\bm e}_{\omega}$ and $\bm E_{2\omega}=E_{2\omega}e^{i\phi_{2\omega}}\hat{\bm
  e}_{2\omega}$. Here $E_{\omega}$ and $E_{2\omega}$ are real and
positive field amplitudes,
$\hat{\bm e}_{\omega}$ and $\hat{\bm e}_{2\omega}$ are 
polarization vectors, and $\phi_{\omega}$ and $\phi_{2\omega}$ are their
phases, and $\Delta \phi\equiv 2\phi_{\omega}-\phi_{2\omega}$ is the
relative-phase parameter that is used to control the current.  In
the following, we give the current injection for different
configurations of the laser beams. 

\subsection{Co-circularly polarized beams}
For two circularly polarized beams propagating along the $z$ direction,
the electric fields are  
$\hat{\bm e}_{2\omega} = \hat{\bm\sigma}^{s_{2}}$ and $\hat{\bm
  e}_{\omega} = \hat{\bm\sigma}^{s_{1}}$ with $s_i=\pm$ identifying for
the handedness and ${\bm\sigma}^s=(\hat{\bm x}+is\hat{\bm y})/\sqrt{2}$. The indirect gap current injection coefficients
are
\begin{eqnarray}
  \dot{\bm J}_{e(h)} &=&
  s_1\text{Im}[\eta_{e(h)}^{xxxx}-\eta_{e(h)}^{xyyx}+2s_1s_2\eta_{e(h)}^{xxyy}]\frac{E^2_{\omega}E_{2\omega}}{\sqrt{2}}\hat{\bm
    m}_{s_1}\ ,\nonumber\\
  \dot{K}_{e(h)}^{ab}&=&[\mu_{e(h)}^{zxyyy} - \mu_{e(h)}^{zxyzz} 
  +s_1s_2 \mu_{e(h)}^{zxzyz}]\frac{E^2_{\omega}E_{2\omega}}{\sqrt{2}}\hat{\bm
    m}_{s_1}^a\hat{\bm z}^b\nonumber\\
  &&\hspace{-1cm}-[\mu_{e(h)}^{zxyyy} - \mu_{e(h)}^{zxyxx} 
  +s_1s_2 \mu_{e(h)}^{zxxxy}]\frac{E^2_{\omega}E_{2\omega}}{\sqrt{2}}\hat{\bm z}^a\hat{\bm
    m}_{s_1}^b
\end{eqnarray}
with $\bm m_{s_1}=s_1 \hat{\bm x}\sin \Delta\phi  +\hat{\bm y}
\cos\Delta\phi$. Both the direction of the charge
and spin currents and the polarization of the spin current can be
controlled by the relative-phase parameter $\Delta\phi$ and the light
polarization $s_i$. The charge current flows only in the $x-y$
plane, and the calculated $\eta_{e(h)}^{xxyy}$ is negligible. For the opposite
circularly-polarized beams, the $x$ component of the charge current
is kept unchanged, but the $y$ component reverses. The spin current
flows in the $x-y$ plane with spin polarization along $x$-axis, or
flows along $z$-direction with spin polarization along $x/y$ direction. 

\subsection{Cross-linearly polarized beams}
For two $z$ propagating cross-linearly polarized beams, $\bm
E_{\omega}$ along the $\hat{\bm x}$ direction and $\bm E_{2\omega}$
along the $\hat{\bm y}$ direction, the injection
current rates are given as
\begin{eqnarray}
  \dot{\bm J}_{e(h)} &=& 2
  \text{Im}[\eta_{e(h)}^{xxyy}]E^2_{\omega}E_{2\omega}\hat{\bm y}\sin\Delta\phi
  \ ,\\
  \dot{K}_{e(h)}^{ab}&=&2(\mu_{e(h)}^{zxyxx}\hat{\bm z}^a\hat{\bm
    x}^b-\mu_{e(h)}^{zxyzz}\hat{\bm x}^a\hat{\bm z}^b)E^2_{\omega}E_{2\omega}\cos\Delta\phi\ .\nonumber
\end{eqnarray}
In this scenario, the charge current and the spin current are
injected with $\pi/2$ phase difference. Therefore, by tuning the
relative-phase parameter $\Delta\phi$, a pure charge
current or pure spin current can be injected. The charge current flows along the second 
harmonic polarization axis, and its amplitude is determined by
$\eta^{xxyy}$, which is zero under the parabolic band
approximation. In our full band structure calculation, it is nonzero
due to the band warping, but very small compared to other tensor
components. The spin current has two components, one involving flow along the $x$
direction with the $z$ spin polarization, and the other involving flow along the
$z$ direction with the $x$ spin polarization.  

\subsection{Co-linearly polarized beams}
For two $z$ propagating beams, both polarized along the $x$ direction, the
injection current rates are given as
\begin{eqnarray}
  \dot{\bm J}_{e(h)} &=& 2
  \text{Im}[\eta_{e(h)}^{xxxx}]E^2_{\omega}E_{2\omega}\hat{\bm x}\sin\Delta\phi
  \ ,\nonumber\\
  \dot{K}_{e(h)}^{ab}&=&2\mu_{e(h)}^{zxyyy}(\hat{\bm y}^a\hat{\bm
    z}^b-\hat{\bm z}^a\hat{\bm y}^b)E^2_{\omega}E_{2\omega}\cos\Delta\phi\ .
\end{eqnarray}
This scenario also gives the phase difference between the charge
current and the spin current as $\pi/2$, so as for cross-linearly
polarized beams pure charge current
injection or the pure spin current injection can also be realized by
choosing a suitable relative-phase parameter $\Delta\phi$. Our results
give the relative-phase parameter dependence of the injected current
as $\sin\Delta\phi$, which is in good agreement with the experimental
results\cite{NaturePhys._3_632_2007_Costa}
around zero probe delay. To
understand the indirect current injection better, we compare the
indirect current injection with the direct one. Because of the lack of
the direct gap injection in silicon in the literature, our results are compared
with the direct current injection in bulk
germanium\cite{Phys.Rev.B_81_155215_2010_Rioux}. 
For the charge currents injected across the indirect gap in silicon, the electron and hole currents have opposite
directions at high photon energies, but they can be the same at low
energies; for charge currents injected across the direct gap in germanium, they always
have the same directions. 
For the spin current, the injected spin current is not so small
compared to other components, 
especially at 300~K, while they are ignorable small in the direct
gap current injection in germanium because of the complete lack of the
helicity of the incident light. 

In this configuration, a good characterization of the charge current
is the swarm velocity, which is defined as the average velocity per injected
carriers forming this current, $v_s^x=\dot{J}^x/e_s\dot{n}$,
with $\dot{n}$ taken from Eq.~(\ref{eq:totaln}). Here $e_s=-e$ is
used for electrons and $e_s=e$ for holes. When $\Delta\phi$ is
a multiple of $\pi/2$ and  the indirect one-photon charge injection
rate equals the indirect two-photon charge injection rate, the
maximum swarm velocity is 
\begin{equation}
  v_{s,\text{max}}^x(\omega) =
  \frac{\text{Im}[\eta_s^{xxxx}(\omega)]}{e_s\sqrt{\xi_s^{xx}(2\omega)\xi_s^{xxxx}(\omega)}}\
  .
\label{eq:swarmv}
\end{equation}
\begin{figure}[htp]
  \centering
  \includegraphics[width=8cm]{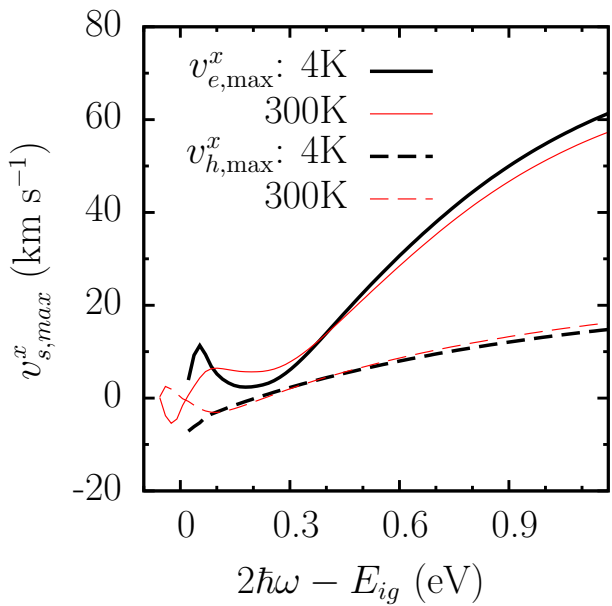}
  \caption{(Color online) Maximal swarm velocity $v^x_{s,max}$ for the
    injected electrons (solid curves) and holes (dashed curves) at 4~K
    (thick black curves) and 300~K (thin red curves).}
\label{fig:swarm}
\end{figure}
\begin{figure}[htp]
  \centering
  \includegraphics[width=8cm]{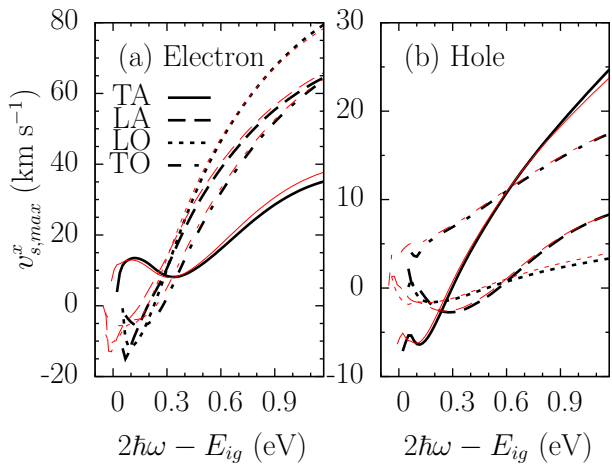}
  \caption{(Color online) Phonon-resolved Maximal swarm velocity for
    the injected (a) electrons and (b) holes at
    4~K (thick black curves) and 300~K (thin red curves).}
\label{fig:swarmphonon}
\end{figure}
We show the maximum swarm velocity in Fig.~\ref{fig:swarm} for the
injected electrons and holes at 4~K and 300~K. The behavior of the swarm
velocity can be divided into two regions: (i) for photon energies in
$2\hbar\omega-E_{ig}\gtrapprox0.25$~eV, the maximum swarm velocities 
are along the $x$ direction for injected electrons and holes, and both
magnitudes increase with increasing photon energy. Compared to the
maximum swarm velocities in bulk germanium, which is in the order of
$10^3$~km/s, the velocity here is about one order of magnitude smaller
due to the larger conduction band effective mass in silicon. (ii) for photon energies in
$2\hbar\omega-E_{ig}<0.2$~eV, the swarm velocities show fine
structures. In particular, all currents experience directional changes
except the electron swarm velocity at 4~K. Analogous to indirect one- and
two-photon charge and spin injection, these fine structures are
induced by the different phonon branches, which are clearly shown in
the phonon-resolved maximal swarm velocity in
Fig.~\ref{fig:swarmphonon}. At 300~K, the injection edge is given by
the TO/LO phonon absorption processes, both of which give negative
velocities for injected electrons and positive velocities for holes around the injection
edge. At 4~K, the injection edge is given by the TA phonon
emission process, which direction is opposite to the band edge current
at 300~K. Therefore, the sign
change of the injected current is induced by the contributions from
different phonon branches. At high photon energies, the injected
velocities are almost independent of the temperature. This is similar
to the temperature dependence of the DSP of the  one-photon indirect
injection\cite{Phys.Rev.B_83_165211_2011_Cheng}: The only temperature dependence in the injection rates
lies in the phonon number, which is the same for the denominator and numerator in
Eq.~(\ref{eq:swarmv}) for a given phonon branch. For high photon
energies, an average phonon number can be used as a good
approximation, and the swarm velocities, given by the ratio
in Eq.~(\ref{eq:swarmv}), are almost temperature independent. 

\section{\label{sec:conclusion}Conclusion}
In conclusion, we have performed a full band structure calculation of
two-photon indirect carrier and spin injection, and
two-color indirect current injection, in bulk silicon. We presented the spectral
dependence for all components of the response tensors at 4~K and 300~K, with which
the injection under any laser beams can be extracted. All injection
rates increase with increasing temperature due to strong
electron phonon interaction at high temperature. We discussed in
detail the injection under different polarized light beams. 

For two-photon indirect optical carrier and spin injections, we considered the injection
under $\sigma^-$ light propagating along $\langle001\rangle$ and
$\langle111\rangle$ directions. For $\langle001\rangle$ light, the
injection rates in the $X$ and $Y$ valleys are the same, but different
from that in the $Z$ valley; for the $\langle111\rangle$ light, the
injections into all valleys are equivalent. For carrier injection,
injections for these two light propagation axes differ slightly. The
calculated injection anisotropy and the linear-circular dichroism
characterize the nonparabolic band effect in the full band structure
calculation. For the $\langle001\rangle$ light, the injection in
the $Z$ valley are much larger than that in the $X$/$Y$ valleys, and
give the valley anisotropy, which is induced by the velocity anisotropy
in the conduction band. At 4~K, the TA phonon-assisted process dominates at
low photon energies, and the TO phonon-assisted process dominates at high photon
energies. At 300~K, the TA phonon-assisted process dominates for all
photon energies. 

For two-photon indirect gap spin injection, the total injected spins orient
parallel to the light propagation direction for the two directions considered. The spin injection rates
increase from the injection edge to a maximum value with the photon energy
increasing, and then decrease. The DSP strongly depends on the
temperature around the injection edge. For the $\langle001\rangle$ light, the injected
spins in each valley are still along the $z$ direction, but the spin
injection rates in the $X$ and $Z$ valleys are different. The maximum
DSP of total spins is $20\%$  at 4~K and $-10\%$ at  300~K; the DSP
can reach about $40\%$ at 4~K and $-20\%$ at 300~K in the $X$ valley, and both become $10\%$ in the $Z$ valleys. For $\langle
111\rangle$ light, the spins in each valley orient to a direction
different from the light propagating direction. In the $Z$ valley, the
$x$ and $y$ (transverse) components have the same injection rates,
which are different from the $z$ (longitudinal)
component. The maximum DSPs of the total spins are $6\%$ at 4~K and
$20\%$ at 300~K, while the one of the $z$ component spin in the $Z$ valley
is about $5\%$ at 4~K and $20\%$ at 300~K, and becomes $25\%$ and
$-8\%$ for the $x$ or $y$ components. All these features are induced
by the interplay of different phonon branch-assisted processes.

For light propagating along the $\langle001\rangle$ direction,
  the injected carriers or spins break the symmetry between
  the $X$ and $Z$ valleys. Such a valley anisotropy of injected
  carriers could be
  probed experimentally, for example, in a pump-probe scenario where the
  probe beam propagates either parallel or perpendicular to the pump
  beam.

For the coherent control, we calculated two-color indirect charge and
spin current injection under three different polarization
configuration of the two-color beams
propagating along the $z$ direction. For the
co-circularly polarized beams, the direction of the injected charge
current is in the $x-y$ plane; the spin current flows in the $x-y$ plane
with a $z$ oriented spin polarization, or flows along the $z$ direction
with the spin orientation in the $x-y$ plane; the current direction or
the spin polarization in the $x-y$ plane can be controlled by a
relative-phase parameter. For the co-linearly polarized beams and the
cross-linearly polarized beams, the directions of the charge current, the spin
current, and the spin polarization are orthogonal to each other. In
these two cases, a pure spin current or a pure charge current can be
obtained by choosing a suitable relative-phase parameter. We
calculated the maximum swarm velocity for charge current as a function
of photon energy, and found 
that the maximum swarm velocities undergo a sign change near the
band edge, which is induced by the contributions from different phonon branches. 

\acknowledgments
This work was supported by the Natural Sciences and Engineering Research
Council of Canada. J.R. acknowledges support from FQRNT.

\appendix
\section{\label{app:injectionrates}Dependence of the injection rates
  on the light propagating direction}
\begin{figure}[htp]
  \centering
  \includegraphics[width=8cm]{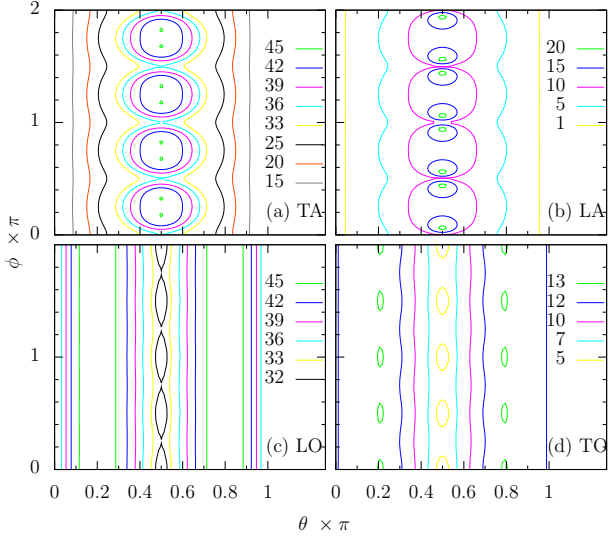}
  \caption{(Color online) Light propagating dependence of the $P_z$ at
    the edge of the each phonon emission process in the $Z$ valley. (a) TA phonon,
    (b) LA phonon, (c) LO phonon, (d) TO phonon. }
  \label{fig:dsptheta}
\end{figure}

For a circularly polarized laser pulse propagating along direction $\hat{\bm
  n}_1$, the electric field can be expressed as
\begin{equation}
  \bm E_{\omega}=\frac{E_0}{\sqrt{2}}(\hat{\bm n}_2 + is\hat{\bm n}_3)
\end{equation}
with
\begin{equation}
\hat{\bm
  n}_1=\begin{pmatrix}\sin\theta\cos\phi\\\sin\theta\sin\phi\\\cos\theta\end{pmatrix}\,,
\quad\hat{\bm n}_2=\begin{pmatrix}\sin\phi\\-\cos\phi\\0\end{pmatrix}\,,
\quad\hat{\bm n}_3=\hat{\bm n}_1\times\hat{\bm n}_2\,,
\end{equation}
here $s=\pm1$ identifies the helicity. In the $Z$ valley the carrier injection rates can be written as  
 \begin{eqnarray}
   \dot{n}_{Z}&=&E_0^4\Big[
    \frac{1}{8}(1+\cos^2\theta)^2(\xi_Z^{xxxx}-\xi_Z^{xxyy}+2\xi_Z^{xyxy})\nonumber\\
    &+&\frac{1}{8}(\sin^2\theta\cos2\phi)^2(\xi_Z^{xxxx}-\xi_Z^{xxyy}-2\xi_Z^{xyxy})\nonumber\\
&+&
\frac{1}{4}\sin^4\theta(\xi_Z^{zzzz}+\xi_Z^{xxyy}-2\xi_Z^{zzxx})\nonumber\\
&+&  \sin^2\theta(1+\cos^2\theta)\xi_Z^{xzxz}\Big]\,,
 \end{eqnarray}
and the spin injection rates as
\begin{eqnarray}
\dot{\bm S}_{Z} &=& E_0^4 s  \Big\{-(\cos^2\theta+1)\cos\theta
\text{Im}[\zeta_Z^{zxyxx}]\hat{\bm z}\nonumber\\
&-& 2\sin^2\theta\cos\theta\text{Im}[\zeta_Z^{zyzxz}] \hat{\bm z}-\sin^3\theta  \text{Im}[\zeta_Z^{xzzyz}]\hat{\bm
  n}_3^{\prime}\nonumber\\
&+& \frac{1}{4}\bm g_1(\theta,\phi) \sin\theta
\text{Im}[-\zeta_Z^{xyzyy}+\zeta_Z^{xyzxx}+2\zeta_Z^{xxzxy}]\nonumber\\
&-&\sin^2\theta\sin\theta \text{Im}[\zeta_Z^{xyzxx}]\hat{\bm
  n}_3^{\prime}\nonumber\\
&-& (3+\cos2\theta)\sin\theta\text{Im}[\zeta_Z^{xxzxy}]\hat{\bm n}_3^{\prime}\Big\}
\end{eqnarray}
with $\hat{\bm n}_3^{\prime}=\hat{\bm z}\times\hat{\bm n}_2$ and 
\begin{equation}
  \bm g_1(\theta,\phi) =
  (\sin^2\theta\sin4\phi)\hat{\bm
    n}_2+[4+\sin^2\theta(\cos4\phi-1) ]
  \hat{\bm n}_3^{\prime}\,.
\end{equation}
For an arbitary propagating direction $(\theta,\phi)$, the direction of the spin
polarization is not always along $\hat{\bm n}_1$. In this case, we
define DSP as the ratio of the magnitude of the spin injection
rate and the carrier injection rate $P_Z=|\dot{\bm
  S}_Z|/(\dot{n}_Z\hbar/2)$. We plot in Fig.~\ref{fig:dsptheta} the $(\theta,\phi)$-dependence of
the $P_z$ at the edge of the each phonon emission process in the $Z$
valley, which shows strong anisotropy of the light propagating
direction. The maximum $P_z$ can reach 45\% at
$(\theta,\phi)\approx(\pi/2,0.2\pi)$ for the TA phonon emission process,
20\% at $\approx(\pi/2,0.1\pi)$ for LA phonon, 45\% at $\theta\approx\pi/4$
for LO phonon, and 13\% at $\theta\approx0.2\pi$ for TO phonon. 

The total carrier injection rates are
\begin{eqnarray}
  \dot{n}&=&E_0^4\xi^{xxxx}\Big[1-\delta-
  \frac{\sigma}{2} \sin^2\theta(\sin^2\phi\cos^2\phi\sin^2\theta+\cos^2\theta)\Big]\,,\nonumber\\
\end{eqnarray}
with $\delta$ and $\sigma$ defined in Eqs.~(\ref{eq:dichroism}). 
The total spin injection rates are
\begin{eqnarray}
\dot{\bm
    S}&=&E_0^4s\Big\{-2 \text{Im}[\zeta^{zxyxx}]\hat{\bm n}_1
  \nonumber\\
&&+\frac{1}{4}\sin\theta \bm
g_2(\theta,\phi)\text{Im}[\zeta^{zxyxx}-2\zeta^{xxzxy}]\Big\}\,,
\end{eqnarray}
with
\begin{eqnarray}
\bm g_2(\theta,\phi) &=&
\sin\theta(1+7\cos^2\theta-\sin^2\theta\cos4\phi )\hat{\bm
  n}_1\nonumber\\
&-&(\sin^2\theta\sin4\phi ) \hat{\bm
  n}_2\nonumber\\
&-&
\cos\theta(3-7\cos^2\theta+\sin^2\theta\cos4\phi)\hat{\bm n}_3\,.
\end{eqnarray}
From above expressions, the two-photon carrier and spin injection rates
show strong anisotropy for the light propagating direction. For the
spin injection, the direction of the injected spin polarization
usually differs from the light propagating direction, but it reverses as
the light helicity changes.

For the linear polarized laser pulse, we also found that the total
carrier injection rates strongly depend on the polarization
direction. For the electric field 
\begin{equation}
  \bm E_{\omega}=E_0(\hat{\bm n}_2\sin\alpha+\hat{\bm n}_3\cos\alpha)\,,
\end{equation}
with $\alpha$ for the polarization direction, the total carrier
injection rates are given by
\begin{eqnarray}
\dot{n}&=&E_0^4\xi^{xxxx}\{1-[1-f(\theta,\phi,\beta)]\sigma\}\,,
\end{eqnarray}
with
\begin{eqnarray}
f(\theta,\phi,\beta) &=&
(\cos\phi\sin\alpha-\cos\theta\sin\phi\cos\alpha)^4\nonumber\\
&+&(\sin\phi\sin\alpha+\cos\theta\cos\phi\cos\alpha)^4\nonumber\\
&+&\sin^4\theta\cos^4\alpha\,.
\end{eqnarray}


\begin{thebibliography}{10}
\bibitem{effectivemass}
  The effective mass along the $i^{th}$ direction in the conduction band can be
  calculated by $\left[m^{\ast}_{c\bm k}\right]^{-1}_{ii}=2\sum_{l\neq
  c}|v^i_{lc\bm k}|^2/(\varepsilon_{l\bm k}-\varepsilon_{c\bm k})+1/m_0$ with
  $m_0$ the free electron mass. In bulk silicon, the longitudinal conduction
  band effective mass is about $0.9$~m$_0$ with m$_0$ being the free electron
  mass, and the transverse one is only $0.2$~m$_0$\cite{landolt_Si}. Thus the
  interband velocity matrix elements along longitudinal direction is
  qualitatively smaller than the transverse one.
\end{thebibliography}
\end{document}